\documentclass[twocolumn, oneside]{aastex631}
\usepackage{etoolbox}
\pretocmd{\abstractname}{\newpage}{}{}

\newcommand{\src}{Swift~J1727}
\usepackage{rotating}
\usepackage{makecell} 

\begin{document}
\maxdeadcycles=1000
\title{Comprehensive Radio Monitoring of the Black Hole X-ray Binary Swift J1727.8$-$1613 during its 2023--2024 Outburst}

\author[0000-0003-0764-0687]{Andrew K. Hughes}
\affiliation{Department of Physics, University of Oxford, Denys Wilkinson Building, Keble Road, Oxford OX1 3RH, UK}


\author[0000-0002-0426-3276]{Francesco Carotenuto}
\affiliation{INAF-Osservatorio Astronomico di Roma, Via Frascati 33, I-00076, Monte Porzio Catone (RM), Italy}

\author[0000-0002-7930-2276]{Thomas D. Russell}
\affiliation{INAF, Istituto di Astrofisica Spaziale e Fisica Cosmica, Via U. La Malfa 153, I-90146 Palermo, Italy}

\author[0000-0003-3906-4354]{Alexandra J. Tetarenko}
\affiliation{Department of Physics and Astronomy, University of Lethbridge, Lethbridge, Alberta, T1K 3M4, Canada}


\author{James C. A. Miller-Jones}
\affiliation{International Centre for Radio Astronomy Research, Curtin University, GPO Box U1987, Perth, WA 6845, Australia}


\author{Arash Bahramian}
\affiliation{International Centre for Radio Astronomy Research, Curtin University, GPO Box U1987, Perth, WA 6845, Australia}

\author[0000-0002-7735-5796]{Joe S. Bright}
\affiliation{Department of Physics, University of Oxford, Denys Wilkinson Building, Keble Road, Oxford OX1 3RH, UK}
\affiliation{Breakthrough Listen, Astrophysics, Department of Physics, The University of Oxford, Keble Road, Oxford, OX1 3RH, UK}

\author[0009-0009-0079-2419]{Fraser J. Cowie}
\affiliation{Department of Physics, University of Oxford, Denys Wilkinson Building, Keble Road, Oxford OX1 3RH, UK}

\author{Rob Fender}
\affiliation{Department of Physics, University of Oxford, Denys Wilkinson Building, Keble Road, Oxford OX1 3RH, UK}
\affiliation{Department of Astronomy, University of Cape Town, Private Bag X3, Rondebosch 7701, South Africa}

\author[0000-0003-0685-3621]{Mark A.~Gurwell}
\affiliation{Center for Astrophysics $|$ Harvard \& Smithsonian, 60 Garden Street, Cambridge, Ma, 02138}

\author{Jasvinderjit K. Khaulsay}
\affiliation{Department of Physics and Astronomy, The University of Sheffield, Hicks Building, Hounsfield Road, Sheffield S3 7RH, UK}

\author{Anastasia Kirby}
\affiliation{International Centre for Radio Astronomy Research, Curtin University, GPO Box U1987, Perth, WA 6845, Australia}

\author{Serena Jones}
\affiliation{International Centre for Radio Astronomy Research, Curtin University, GPO Box U1987, Perth, WA 6845, Australia}

\author{Elodie Lescure}
\affiliation{Department of Physics and Astronomy, University of Lethbridge, Lethbridge, Alberta, T1K 3M4, Canada}

\author[0000-0002-8384-3374]{Michael McCollough}
\affiliation{Center for Astrophysics $|$ Harvard \& Smithsonian, 60 Garden Street, Cambridge, Ma, 02138}

\author{Richard M. Plotkin}
\affiliation{Department of Physics, University of Nevada, Reno, NV 89557, USA}
\affiliation{Nevada Center for Astrophysics, University of Nevada, Las Vegas, NV 89154, USA}

\author[0000-0002-1407-7944]{Ramprasad Rao}
\affiliation{Center for Astrophysics $|$ Harvard \& Smithsonian, 60 Garden Street, Cambridge, Ma, 02138}

\author[0000-0002-7521-9897]{Saeqa D. Vrtilek}
\affiliation{Center for Astrophysics $|$ Harvard \& Smithsonian, 60 Garden Street, Cambridge, Ma, 02138}

\author[0000-0001-7361-0246]{David R. A. Williams-Baldwin}
\affiliation{Jodrell Bank Centre for Astrophysics, School of Physics and Astronomy, The University of Manchester, Manchester, M13 9PL, UK}

\author[0000-0002-2758-0864]{Callan M. Wood}
\affiliation{International Centre for Radio Astronomy Research, Curtin University, GPO Box U1987, Perth, WA 6845, Australia}


\author{Gregory R. Sivakoff}
\affiliation{Department of Physics, University of Alberta, CCIS 4-181, Edmonton, AB T6G 2E1}


\author[0000-0002-3422-0074]{Diego Altamirano}
\affiliation{School of Physics \& Astronomy, University of Southampton, SO17 1BJ, UK}

\author[0000-0002-0752-3301]{Piergiorgio Casella}
\affiliation{INAF-Osservatorio Astronomico di Roma, Via Frascati 33, I-00076, Monte Porzio Catone (RM), Italy}

\author[0000-0001-5538-5831]{Stéphane Corbel}
\affiliation{Université Paris Cité and Université Paris Saclay, CEA, CNRS, AIM, F-91190 Gif-sur-Yvette, France}

\author[0000-0003-3197-2294]{David R. DeBoer}
\affiliation{Radio Astronomy Lab, University of California, Berkeley, CA, USA}

\author[0000-0002-1793-1050]{Melania Del Santo}
\affiliation{INAF, Istituto di Astrofisica Spaziale e Fisica Cosmica, Via U. La Malfa 153, I-90146 Palermo, Italy}

\author[0000-0001-8436-1847]{Constanza Echibur\'u-Trujillo}
\affiliation{Department of Astrophysical and Planetary Sciences, JILA, Duane Physics Bldg., 2000 Colorado Ave., University of Colorado, Boulder, CO 80309, USA}

\author[0000-0002-0161-7243]{Wael Farah}
\affiliation{SETI Institute, 339 Bernardo Ave, Suite 200 Mountain View, CA 94043, USA}

\author[0000-0002-1793-1050]{Poshak Gandhi}
\affiliation{School of Physics \& Astronomy, University of Southampton, SO17 1BJ, UK}

\author[0000-0002-9677-1533]{Karri I. I. Koljonen}
\affiliation{Department of Physics, Norwegian University of Science and Technology, NO-7491 Trondheim, Norway}

\author{Thomas Maccarone}
\affiliation{Department of Physics and Astronomy, Texas Tech University, Lubbock, TX 79409-1051, USA}

\author[0000-0002-3493-7737]{James H. Matthews}
\affiliation{Department of Physics, University of Oxford, Denys Wilkinson Building, Keble Road, Oxford OX1 3RH, UK}

\author[0000-0001-9564-0876]{Sera B. Markoff}
\affiliation{Anton Pannekoek Institute for Astronomy, University of Amsterdam, Science Park 904, 1098 XH Amsterdam, The Netherlands}
\affiliation{Gravitation and Astroparticle Physics Amsterdam Institute, University of Amsterdam, Science Park 904, 1098 XH 195 196 Amsterdam, The Netherlands}

\author[0000-0002-3430-7671]{Alexander W. Pollak}
\affiliation{SETI Institute, 339 Bernardo Ave, Suite 200 Mountain View, CA 94043, USA}

\author[0000-0002-3500-631X]{David M. Russell}
\affiliation{Center for Astrophysics and Space Science (CASS), New York University Abu Dhabi, PO Box 129188, Abu Dhabi, UAE}

\author[0000-0002-5319-6620]{Payaswini Saikia}
\affiliation{Center for Astrophysics and Space Science (CASS), New York University Abu Dhabi, PO Box 129188, Abu Dhabi, UAE}

\author[0000-0002-5870-0443]{Noel Castro Segura}
\affiliation{Department of Physics, University of Warwick, Gibbet Hill Road, Coventry, CV4 7AL, UK}

\author[0000-0002-8808-520X]{Aarran W. Shaw}
\affiliation{Department of Physics and Astronomy, Butler University, 4600 Sunset Avenue, Indianapolis, IN 46208, USA}

\author[0000-0003-2828-7720]{Andrew Siemion}
\affiliation{Department of Physics, University of Oxford, Denys Wilkinson Building, Keble Road, Oxford OX1 3RH, UK}
\affiliation{Breakthrough Listen, Astrophysics, Department of Physics, The University of Oxford, Keble Road, Oxford, OX1 3RH, UK}
\affiliation{SETI Institute, 339 Bernardo Ave, Suite 200, Mountain View, CA 94043, USA}
\affiliation{Berkeley SETI Research Centre, University of California, Berkeley, CA 94720, USA}
\affiliation{Department of Physics and Astronomy, University of Manchester, UK}
\affiliation{University of Malta, Institute of Space Sciences and Astronomy, Msida, MSD2080, Malta}

\author[0000-0002-4622-796X]{Roberto Soria}
\affiliation{INAF-Osservatorio Astrofisico di Torino, Strada Osservatorio 20, I-10025 Pino Torinese, Italy}
\affiliation{Sydney Institute for Astronomy, School of Physics A28, The University of Sydney, Sydney, NSW 2006, Australia}

\author{John A. Tomsick}
\affiliation{Space Sciences Lab, University of California, Berkeley, 7 Gauss Way, Berkeley, CA 94720, USA}

\author[0000-0002-5686-0611]{Jakob van den Eijnden}
\affiliation{Department of Physics, University of Warwick, Coventry CV4 7AL, UK}
\affiliation{Anton Pannekoek Institute for Astronomy, University of Amsterdam, Science Park 904, 1098 XH Amsterdam, The Netherlands}




\begin{abstract}
This work presents comprehensive multi-frequency radio monitoring of the black hole low-mass X-ray binary Swift J1727.8$-$1613, which underwent its first recorded outburst after its discovery in August 2023. Through a considerable community effort, we have coalesced the data from multiple, distinct observing programs; the light curves include ${\sim}\,10$\,months and 197 epochs of monitoring from 7 radio facilities with observing frequencies ranging from (approximately) 0.3--230\,GHz. The primary purpose of this work is to provide the broader astronomical community with these light curves to assist with the interpretation of other observing campaigns, particularly non-radio observing frequencies. We discuss the phenomenological evolution of the source, which included: (i) multiple radio flares consistent with the launching of discrete jet ejections, the brightest of which reached $\sim$\,1\,Jy; (ii) temporally evolving radio spectral indices ($\alpha$), reaching values steeper than expected for optically-thin synchrotron emission ($\alpha\,{<}\,-1$) and emission with significant radiative cooling ($\alpha\,{<}\,-1.5$). We have published a digital copy of the data and intend for this work to set a precedent for the community to continue releasing comprehensive radio light curves of future low-mass X-ray binary outbursts.\end{abstract}



\section{Introduction} \label{sec:intro}

Astrophysical jets are highly collimated outflows that are produced by nearly all accreting compact objects \citep[black holes, neutron stars, and white dwarfs;][]{livio2002,kumar2015,ruiz2021,Fender2019sscyg,DeColle2020,Gottlieb2023}. Given their ubiquity, interactions between the matter accelerated by jets and the ambient environment (i.e., `feedback') have played and continue to play a critical role in driving local evolution on even the largest of scales \citep[e.g.,][]{Hardcastle2020}. Therefore, understanding jets is crucial for understanding the Universe as a whole. For time-domain studies, an archetypal jet `laboratory' is the black hole low-mass X-ray binary (BH LMXB). 

Black hole low-mass X-ray binaries are interacting binaries composed of a stellar-mass black hole (${\sim}\,10{\rm\,M_\odot}$) accreting material from a low-mass (${\lesssim}\,1{\rm\,M_\odot}$) companion star. Many BH LMXBs are transient in nature, spending the majority of their lifetimes accreting small amounts of matter in a quiescent low-luminosity state \citep[$L_X\,{<}\,10^{34}{\rm\,erg\,s^{-1}}$, where $L_X$ is the X-ray luminosity of the accretion flow;][]{Plotkin2013QS}. These objects may also sporadically enter into bright outbursts where their broadband emission brightens considerably. During the week-to-years-long outbursts \citep{btetarenko2016,BlackCAT}, the accretion flow \citep[best observed at X-ray frequencies, e.g.,][]{Tananbaum1972,Belloni1999} and jets \citep[best observed at radio-to-infrared frequencies, e.g.,][]{Tananbaum1972,corbel2002,russel2013,atetarenko2017} evolve significantly as the system transitions through different empirically defined `accretion states' \citep[see, e.g.,][]{fender2004,mclintock2006,belloni2010}.

A typical outburst begins with a rapid increase in the X-ray luminosity and a transition (from quiescence) into the \textit{hard state} ($10^{35}\,{\lesssim}\,L_X\,{\lesssim}\,10^{39}{\rm\,erg\,s^{-1}}$). In the hard state, the X-ray spectrum is non-thermal, with the X-rays being dominated by high energy emission (i.e., `hard' X-rays) due to Compton up-scattering of seed photons by an optically thin coronal flow \citep{Thorne1975,Shapiro1976}. As the accretion rate increases, the source transitions through intermediate states: first, the \textit{hard intermediate state} followed by the \textit{soft intermediate state}, during which the X-ray emission becomes progressively softer (i.e., more dominated by lower energy `soft' disc photons). Eventually, the system enters the \textit{soft state}, with a spectrum dominated by thermal emission from an optically thick, geometrically thin accretion disc. A BH LMXB typically remains in the soft state for weeks to months, decreasing in X-ray luminosity before transitioning back to the hard state through lower-luminosity intermediate states and finally returning to quiescence. \citep[see, e.g.,][for a comprehensive description of the accretion states]{Jhoman2005,mclintock2006,belloni2010,ingram2019,Kalemci2022review}.

The properties of relativistic jets evolve with the accretion states \citep[see, e.g.,][for detailed reviews]{fender2004,fender2009,fender2010}. In the hard state, a continuously accelerated jet is launched (henceforth the `compact jet'), typically described with the stratified, conical outflow model of \citet{blandfordkonigl1979}. Compact jets exhibit an optically thick, partially self-absorbed synchrotron spectrum with a (typically) flat or mildly inverted spectral index ($\alpha\,{\gtrsim}\,0$; the flux density, $F_{R,\nu}$ is proportional to the observing frequency, $\nu$, such that $F_{R,\nu} \propto \nu^\alpha$) up to a break frequency where the spectrum becomes optically thin \citep[$\alpha\,{\sim}\,-0.6$, typically in the sub-mm or near-infrared regimes; ][]{migliari2010,russel2013,russell2014}.  The radio luminosity from the compact jet shows a strong, positive correlation with X-ray luminosity from the accretion flow \citep[e.g.,][]{gallo2003,corbel2013,gallo2018}, consistent with continuous accretion-jet coupling in the hard state. 

Monitoring campaigns following state transitions have captured ${\geq}\,3\,$orders of magnitude decreases in radio luminosity, showing that compact jets are strongly quenched in the soft state \citep[e.g.,][]{Fender1999gx339quench,coriat2011,bright2020,russell2019,russel2020,Maccarone2020quench,Carotenuto2021}. Around this time, bright radio flares can be observed around the hard-to-soft state transition \citep[e.g.,][]{atetarenko2017}. These flares are associated with the launching of mild-to-extremely relativistically bipolar ejecta, decoupling the jets from the accretion flow \cite[e.g.,][]{mirabel1994,Hjellming1995,Fender1999,Corbel2002Ejecta,Brocksopp2002ejection,rushton2017,millerjones2019,russel2019,bright2020,Carotenuto2021,wood2021}. The radio spectra of the flares evolve rapidly in time, transitioning from optically thick to optically thin synchrotron emission as the emitting plasma expands, causing a decrease in the self-absorption frequency \citep[e.g.,][]{vdl1966,Hjellming1988,fender2019}. As a result, for the majority of their lifetimes (including when they are spatially resolvable), ejecta are optically thin radio sources.

Jet ejecta can be detected from minutes to even years after launch \citep[e.g.,][]{Corbel2002Ejecta,millerjones2019,Bahramian2023,wood2021}. An observational consequence is that radio emission can be detected in the soft state despite the compact jet being quenched. The longest-lasting ejecta are believed to remain bright because of the continual acceleration of particles driven by interactions with the ambient interstellar medium \citep[ISM;][]{Corbel2002Ejecta,russell2019,Espinasse2020,bright2020,Carotenuto2021}, where some BH LMXB jets produce interaction regions tens of parsecs away from the BH \citep[e.g., ][]{Tetarenko2018b}. In extreme cases, ejecta that have faded below the detection threshold can remain undetected for months or years before re-brightening during jet-ISM interactions \citep[e.g.,][]{Corbel2002Ejecta,Carotenuto2021}. 

While there is some phenomenological understanding of the evolution of the jets through the BH LMXB accretion states, knowledge of the precise timings and causal sequence of events leading to jet quenching and transient jet launching, or the intrinsic properties such as total energy content (and thus, total energy available for feedback) is far from complete. Each subsequent BH LMXB outburst provides (often multiple) new opportunities for real-time monitoring of jet launching, disruption, and interaction events and, thus, a chance to test and refine our current understanding. Moreover, multi-wavelength observations allow us to probe the connection between jets and the accretion flow, jets and other outflows \citep[e.g., winds;][]{Sanchez2020maxi1820winds} or investigate how jet properties change with the observing frequency \citep{Russell2006LoLx}. Naturally, bright outbursts provide the highest-quality monitoring, but only if the observing campaigns are sufficient to build a coherent picture; one such opportunity arose in 2023 when the BH LMXB Swift J1727.8$-$1613 began a bright outburst.

\subsection{The 2023--2024 Outburst of Swift J1727.8$-$1613}
Swift J1727.8$-$1613 (hereafter \src) was discovered on 2023 Aug 24 (MJD 60180) by the burst alert telescope \citep[BAT;][]{BAT2005,BAT2013} aboard the Neil Gehrels \textit{Swift} Observatory \citep{Swift2004}. Within days, the source's X-ray flux reached ${\sim}\,7\,$Crab (at 2--20\,keV), resulting in its initial classification as a gamma-ray burst \cite[GRB 230824A;][]{page2023}. Follow-up X-ray observations refuted a GRB origin, reclassifying (and renaming) the source as the Galactic transient Swift J1727.8$-$1613 \citep[hereafter Swift J1727,][]{Kennea2023}. The rapid rise in X-ray luminosity motivated extensive multi-wavelength monitoring and identified Swift J1727 as a BH LMXB \citep{Mata2025distance1727}.

Rapid radio follow-up localized the source with sub-arcsecond precision (J2000), $17$h$27$m$43.31(\,{\pm}\,0.04$s), $-16^\circ12^\prime19.23({\pm}\,0.02^{\prime\prime}$) \citep{millerjones2023a}. Multi-epoch optical spectroscopy by \cite{Mata2024distance1727} estimated the properties of the companion star, favouring an early K-type companion star with an ${\sim}\,8\,$hr orbital period. \cite{Mata2024distance1727} used the orbital period and observed optical extinction to estimate an initial distance of $2.7\,{\pm}\,0.3\,$kpc; \citet{Mata2025distance1727} refined the orbital period, calculated a new distance of $3.7\,{\pm}\,0.3\,$kpc, and dynamically confirmed that \src\ contained a stellar-mass black hole. The most recent analysis of optical (and near-UV) extinction by \citet{Burridge2025distance1727} reviewed the systematics in \citet{Mata2025distance1727} and favoured a larger distance of $5.5_{-1.1}^{+1.4}\,$kpc. X-ray spectral modelling revealed relativistically broadened iron lines, suggestive of a high-spin ($a\,{\sim}\,0.98$), medium-inclination system \citep[$\sim$\,40--50$^\circ$ inclination angle;][]{Draghis2023,Peng2024}. Subsequent modelling, which also considered the X-ray polarisation properties, found a similar inclination angle ($\sim$\,30--50$^\circ$) but favoured a lower spin \citep[$a\,{\sim}\,0.87$;][]{Svoboda2024}.

The initial rise in X-ray luminosity occurred as Swift J1727 exhibited X-ray and radio properties characteristic of the hard accretion state \citep[e.g.,][]{Bollemeijer2023a,millerjones2023a,Mereminskiy2024, Peng2024,Liu2024}; direct imaging of the compact jet revealed Swift J1727 to have the largest resolved compact jet seen from a BH LMXB \citep{Wood2024}. As Swift J1727 evolved (through the hard intermediate state), X-ray monitoring suggested the source transitioned to soft-intermediate or soft state on 2024 October 5 \citep[MJD 60222, although X-ray observations were unable to discriminate between the states;][]{Bollemeijer2023a,Bollemeijer2023b}. Radio quenching accompanied the X-ray transition, followed by bright (radio) flaring, consistent with compact jet disruption and the launching of ejecta \citep{millerjones2023b}. Subsequent X-ray observations, noting variability in the hardness ratio, were suggestive of multiple returns to the intermediate states \citep[e.g.,][]{Wenfei2023}. Swift J1727 took ${\sim}\,6\,$months to transition back to the hard state and, by that time, its X-ray luminosity decreased by more than two orders of magnitude \citep[on $\sim\,$MJD 60385, 2024 March 15;][]{Podgorny2024,Russel2024}. \src\ has since returned to quiescence. 

This paper combines observations from multiple collaborations to present a multi-wavelength, multi-facility radio light curve of Swift J1727 during its 2023--2024 outburst. In addition to flux densities, we include, when available, interband radio spectral indices but omit all other properties (e.g., morphology, proper motion, or polarization), as the comprising collaborations have identified them as the focus of \textit{in prep} publications. Our goal is to provide the broader community with comprehensive radio light curves to aid with interpreting the evolution of Swift J1727, particularly for observing programmes at other wavelengths (e.g., X$-$ray or optical monitoring). The rest of the paper is structured as follows: Section \ref{sec:methods} describes the observing and analysis strategies for each instrument; Section~\ref{sec:results} presents the light curves and a qualitatively focused interpretation of the source evolution; finally, Section~\ref{sec:conclusions} presents a summary of the observations. 

\section{Data and Analysis} \label{sec:methods}
\begin{turnpage}
    \begin{figure}[htp]
        \centering
            \includegraphics[scale=0.55]{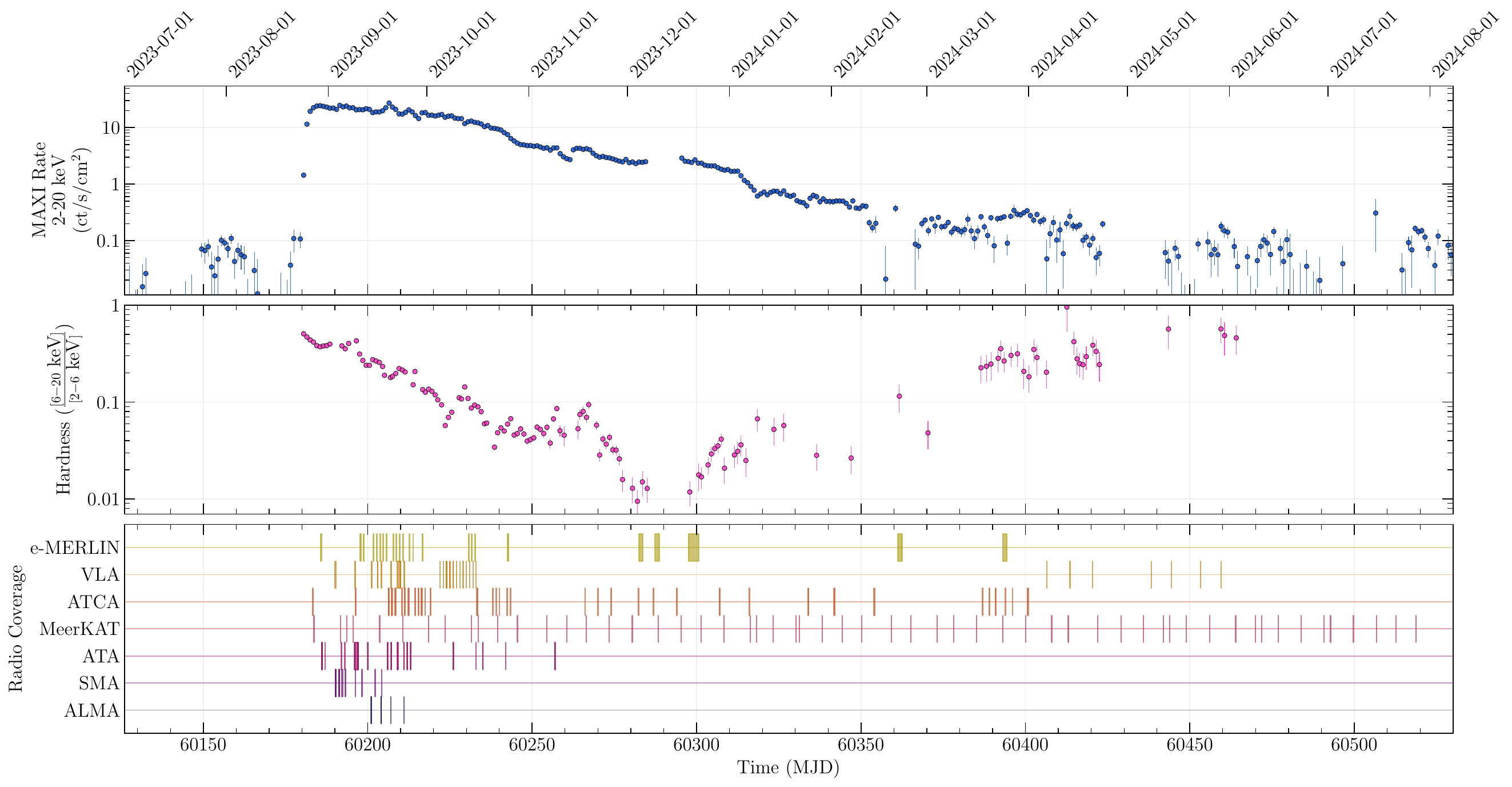}
            \caption{Representation of the radio coverage during the 2023--2024 monitoring campaign of Swift J1727. The MAXI/GSC 2--20\,keV rate (\textit{top panel}); The hardness ratio calculated with using the 2--6\,keV and 6--20\,keV sub-bands (\textit{middle panel}); The coverage from each radio facility (\textit{bottom panel}). When calculating the hardness ratio, we average together adjacent observations until both bands are detected at 3$\sigma$; as a result, the top and middle panels have different time samplings. In the bottom panel, each instrument is represented by a different colour, and the width of each line is the duration of each observation.}
            \label{fig:radio_coverage}
    \end{figure}
\end{turnpage}

This section summarizes each instrument's data reduction, calibration, imaging, and analysis techniques. We focus on concise summaries as the collaborations will publish stand-alone papers with more comprehensive descriptions of their observation and analysis routine(s). In Figure~\ref{fig:radio_coverage}, we include a graphical representation of the radio coverage.

\subsection{MeerKAT}
\label{sec:methods_mkat}
\src\ was observed with the radio interferometer MeerKAT \citep{MeerKATCitation} as a part of the ThunderKAT \citep{tkat} and X-KAT programmes (MeerKAT Proposal ID: SCI-20230907-RF-01). MeerKAT is a 64-element interferometer with a maximum baseline length of 8\,km. The MeerKAT monitoring includes 52 epochs taken on an approximately weekly observing cadence. Most observations (i.e., 47 epochs) used the L-band receivers ($\sim$\,1.28\,GHz; 856\,MHz un-flagged bandwidth). Each L-band observation consisted of a single scan of either 15 or 30 minutes on-source, flanked by two 2-minute scans of a nearby gain calibrator (J1733$-$1304). Two 5-minute scans of J1939$-$6342 (PKS B1934$-$638) were included at the beginning and end of each observation for bandpass and flux scale calibration. In addition to the weekly L-band monitoring, five epochs were acquired using MeerKAT's S-band receivers. One of the S-band observations used the S2 ($\sim$\,2.62\,GHz; 875\,MHz un-flagged bandwidth) sub-band, and the other four used the S4 ($\sim$\,3.06\,GHz; 875\,MHz un-flagged bandwidth) sub-band. The S-band monitoring followed the L-band observing strategy.

The data were flagged, calibrated, and imaged using \textsc{polkat}\footnote{\url{https://github.com/AKHughes1994/polkat}} \citep{polkatascl}, which is based on the semi-automated routine \textsc{oxkat} \citep{oxkat2020}, but modified to handle full polarisation observations. This work only includes total intensity observations (i.e., Stokes $I$), and thus, \textsc{oxkat} and \textsc{polkat} have near-identical workflows. Readers are directed to \citet{mightee} for a comprehensive description of the workflow but should note that calibration stopped after the direction-independent self-calibration step (2GC in \textsc{oxkat} parlance). To optimize for high-sensitivity imaging, the image weighting adopted a Briggs' robustness of 0 \citep{Briggs1995}\footnote{MeerKAT's synthesized beam becomes non-Gaussian for robustness weightings that would typically be used when maximizing for sensitivity (i.e., ${>}\,0$); non-Gaussian beam shapes can inhibit accurate deconvolution, thereby raising the image-plane noise.}.

The source properties were measured using the Common Astronomy Software Applications package \citep[\textsc{casa}v6.4;][]{Casa2022} task \textsc{imfit}, fitting elliptical Gaussian component(s) in a small sub-region 

\onecolumngrid
\newpage
\twocolumngrid

\noindent around the source. Multiple Gaussians were simultaneously fit for observations with multiple components (e.g., due to jet ejecta). As the components were unresolved point sources, the component shapes were fixed to the synthesized beam of each image. The (1$\sigma$) uncertainty on the flux densities was measured from the local root-mean-square (rms) noise extracted from an emission-free region centred near the source(s); the extraction region had an area equal to ${\sim}\,$100 PSFs. No further data manipulation was applied.

\subsection{Enhanced Multi-Element Remote-Linked Interferometer Network (e-MERLIN)}

\src\ was also monitored with the \textit{e}-MERLIN radio-interferometer \citep{Garrington_2016} for 24 epochs in total, from August 2023 to March 2024 (project codes RR16003 and CY16208). \textit{e}-MERLIN is a 7-antenna array with a maximum baseline of 217\,km. The first part of the monitoring consisted of a dense coverage at the beginning of the outburst, with 19 observations performed almost daily in September 2024 and during the second half of October 2024, with durations that ranged from 1 to 5.5 hours per epoch, depending on the expected flux density of the target. In this period, observations were performed at C-band, with a central frequency of 5.07 GHz, and total bandwidth of 512 MHz.

The second part of the monitoring started on 2023 December 4 (MJD 60283) and ended on 2024 March 24 (MJD 60393), with five observations log-spaced in time and durations of up to 10 hours. In this part of the campaign, observations were performed at L-band, at a central frequency of 1.51 GHz and total bandwidth of 512 MHz.

For the entire campaign, 3C286 and OQ208 were observed as the flux and bandpass calibrators, respectively, while J1724--1443 was used as the gain calibrator. The data were processed using the \textit{e}-MERLIN CASA Pipeline\footnote{\url{https://github.com/e-merlin/eMERLIN_CASA_pipeline}} \citep{eMCP} version v1.1.19. The pipeline handles the automated \textit{a priori} flagging with \textsc{aoflagger} \citep{Offringa_AOFLAGGER} and the standard bandpass and gain calibration using \textsc{casa}. For each epoch, since it was intended to build a homogeneous light curve from epochs with very different samplings of the \textit{uv}-plane, the flux density of \src\ was extracted through visibility-plane fitting, using \textsc{uvmultifit} \citep[assuming a point source;][]{uvmultifit}.


\subsection{Allen Telescope Array (ATA)}
Observations of Swift J1727 with the Allen Telescope Array (ATA; \cite{bright2023}; Pollack et al. \textit{in prep,}; Farah et al. \textit{in prep,}) began on 2023 Aug 30 with regular monitoring occurring until 2023 Nov 08. These observations used 20 (out of 42) of the ATA antenna and had a maximum baseline length of 300\,m. The ATA is equipped with wide bandwidth feeds, sensitive to frequencies between ${\sim}\,1$ and ${\sim}\,10\,\rm{GHz}$. The current digitisation capabilities of the ATA backend allow for two ${\sim}\,700\,\rm{MHz}$ bandwidth tunings to be placed independently within this frequency range. The observing campaign included data at central frequencies of $1.5$, $3$, $5$, and $8\,\rm{GHz}$. The flux/bandpass and gain calibrators were 3C286 and J1733$-$1304, respectively. The gain calibrator was visited every 30 minutes at all frequencies. Following the standard procedures, data were reduced in \textsc{casa}; imaging was performed using \textsc{wsclean}; the flux density of Swift J1727 was measured following Section~\ref{sec:methods_mkat}.

\subsection{Karl G.~Jansky Very Large Array (VLA)}
 This work includes three separate observing programs with the VLA. We combine data taken in multiple VLA configurations, and, as a result, the 27-antenna array has a (program-dependent) maximum baseline of 3--36\,km. When available, we also include ${\sim}\,$0.3\,GHz flux densities taken with The VLA Low-band Ionosphere and Transient Experiment \citep[VLITE; e.g.,][]{Peters2023VliteJ1727}. VLITE is a commensal instrument on the VLA operated by The U.S. Naval Research Laboratory (NRL); a description of the VLITE processing pipeline can be found in \citet{Polisensky2016VLITE}.

The most rapid response observations presented in this work were taken on 2023 August 25 \citep[MJD 60181;][]{millerjones2023a} as a part of the JACPOT program (Project Code: VLA/23A--260). Initially approved for rapid response C-band (4--8 GHz) observations, JACPOT also monitored the source daily at C- and X-band (8--12 GHz) during the bright flaring seen in October 2023. The rapid response observations were taken using array's most extended A-configuration, whereas the flaring occurred when the source was in a hybrid configuration (A$\rightarrow$D). Each JACPOT observation was 20--30 minutes long and used the VLA's 8-bit samplers, which break each band into two base-bands with 1.024 GHz of bandwidth.

Additionally, during the rising hard state of the outburst, \src\ was also observed by the PITCH-BLACK collaboration (Project Code: VLA/22B--069). The multi-hour PITCH-BLACK observations split the 27-element array (in A-configuration) into three sub-arrays of 10, 9, and 8 antennas. Observations in each sub-array were also made with the 8-bit samplers, and each sub-array observed exclusively in one frequency band: C (4--8 GHz), X (8--12 GHz), or Ku (12--14 GHz) band 

Following the source's return to the hard state, the VLA observed \src\ nine additional times on an approximately weekly cadence (Project Code: VLA/23B--064). Observation durations ranged from 10 to 60 minutes, where longer epochs were taken as the source faded. Observations were recorded at C-band with 3-bit samplers, resulting in 4.096\ GHz bandwidth centred at 6.2\ GHz. The epochs of this final VLA program were evenly split between B- and C-configuration.

All three VLA projects used 3C286 (J1331$+$305) for bandpass and flux scale calibration and J1733$-$1304 for gain calibration. The multi-hour, sub-arrayed observations (i.e., VLA/22B--069) implemented a custom non-periodic target/calibrator cycle for each sub-array, alternating between \src\ and the calibrators at one band per sub-array, such that simultaneous data in all three bands was obtained. The data were reduced and imaged within \textsc{casa}, with flagging and calibration performed using the VLA \textsc{casa} pipeline\footnote{The VLA programs used \textsc{casa} version 6.2, 6.2, and 6.3 for project codes VLA/23A--260, VLA/22B--069, and Project code VLA/23B--064, respectively}. The images were weighted with natural or robust weighting (robustness parameter 0). When necessary, phase-only self-calibration was used to improve image fidelity. Flux densities were measured by fitting a point source in the image plane, as outlined in Section~\ref{sec:methods_mkat}.  

\subsection{Australia Telescope Compact Array (ATCA)}
The Australia Telescope Compact Array (ATCA) observed \src\ thirty-nine times during its 2023--2024 outburst under project codes C2601, C3057, C3362. ATCA is an East-West array with six antennas and a maximum baseline length of 6\,km. For all observations, data were recorded simultaneously at central frequencies of 5.5\,GHz and 9\,GHz, with each frequency band having 2.048\,GHz of un-flagged bandwidth. Observations used either J1939$-$6342 (PKS~B1934$-$638) or PKS~B0823$-$500 for bandpass and flux density calibration depending on source visibility at the time of the observation; J1939$-$6342 was the preferred calibrator. The nearby (${\sim}\,3.4^{\circ}$ away) source J1733$-$1304 (PKS~B1730$-$130) was used for gain calibration. Data were first edited for RFI, before being calibrated and imaged following standard procedures within \textsc{casa}v5.1.2. Imaging typically used a Briggs robust parameter of 0, balancing sensitivity and resolution. However, at later times, when the source was faint, more natural weightings (robustness parameter ${>}\,0$) were used to increase sensitivity. Data were phase-only self-calibrated when the source flux density exceeded $\sim$\,10\,mJy. The flux density of the source was measured following the same procedure as Section~\ref{sec:methods_mkat}.

\subsection{Atacama Large Millimeter Array (ALMA)}
Alongside the PITCH-BLACK observations taken with the VLA, \src\ was observed with ALMA (Project Code: 2022.1.01182.T) four times in September 2023. Data were taken in Band 3, Band 4, and Band 6 at central frequencies of 97.5 GHz, 145.0 GHz, and 233 GHz, respectively. The ALMA correlator was set up to yield 4$\times$2 GHz wide base-bands. During the observations, the array was in its Cycle 10 C8 configuration, with 42--46 antennas and a maximum baseline length of 8.5\,km. The data were reduced and imaged within \textsc{casa}, with flagging and calibration performed using the ALMA \textsc{casa}v6.2 pipeline. J1924$-$2914 was used as a flux/bandpass calibrator, J1733$-$1304 was used as a phase calibrator, and J1742$-$1517 was used as a check source in all bands. Phase-only self-calibration was performed, and imaging followed standard procedures with a natural weighting scheme to maximize sensitivity. Flux densities of the source were then measured by fitting a point source in the image plane (with the \textsc{imfit} task).

\subsection{Submillimeter Array (SMA)}
\src\ was observed with the SMA eight times in September 2023. The SMA correlator yields 12 GHz of bandwidth per sideband per polarization and was tuned to a frequency of 225.5 GHz. 
During the observations, the array was in the \textit{subcompact} configuration, with seven available antennas and a maximum baseline length of 25\,m. The data were reduced and imaged with multiple reduction paths using in-house pipeline routines, the in-house reduction suite \textsc{mir}\footnote{\url{https://lweb.cfa.harvard.edu/rtdc/SMAdata/process/mir/}}, and \textsc{miriad} \citep{miriad1995}, assuring a consistent outcome. 
The flux density scale was set using MWC349A, BL Lac was the bandpass calibrator, and NRAO 530 (J1733$-$1304) was used as the gain calibrator. The target is a point source at the spatial resolution of the SMA during the observations ($\sim$\,5 arcsec), and source properties were extracted through visibility-plane fitting.

\begin{figure*}[t]
    \centering
    \includegraphics[width=1.0\linewidth]{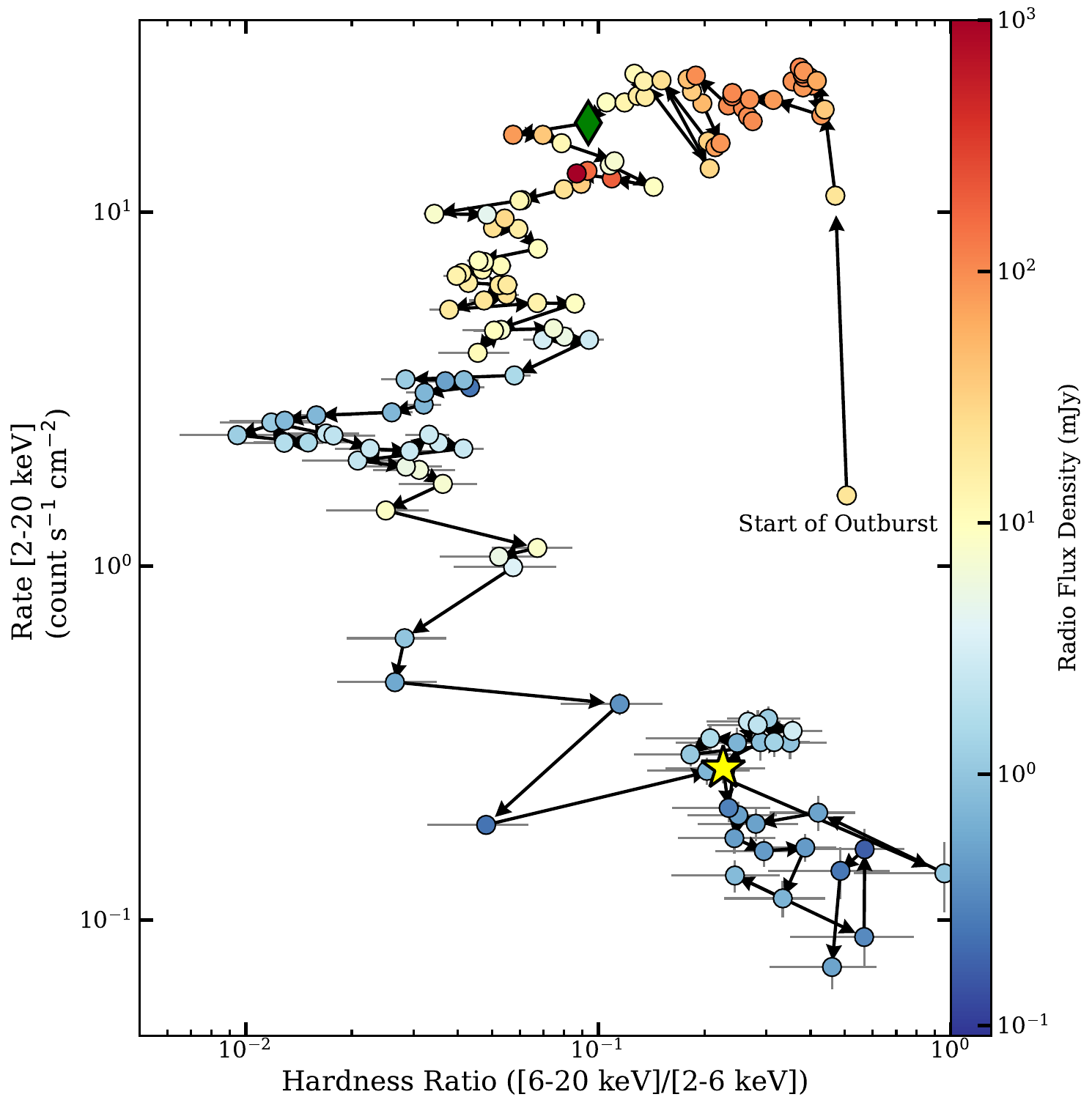}
    \caption{X-ray hardness intensity diagram of Swift J1727 when it was detectable by MAXI/GSC. The marker colours correspond to the scaled 10\,GHz radio flux densities, linearly interpolated (in time) to the MAXI/GSC observing epochs. The colour bar has been capped at 1\,Jy to show the radio evolution more clearly. We highlight the observations where the source was first identified to undergo a hard-to-soft \citep[green diamond, MJD 60222;][]{Bollemeijer2023a,Bollemeijer2023b} and a soft-to-hard \citep[yellow star, MJD 60385;][]{Podgorny2024} state transition. From the HID, it is clear that Swift J1727 has completed the `loop', marking the end of the source's X-ray activity.}
    \label{fig:XrayHID}
\end{figure*}

\onecolumngrid
\newpage
\begin{turnpage}
    \begin{figure}[h]
        \centering
            \includegraphics[scale=0.56]{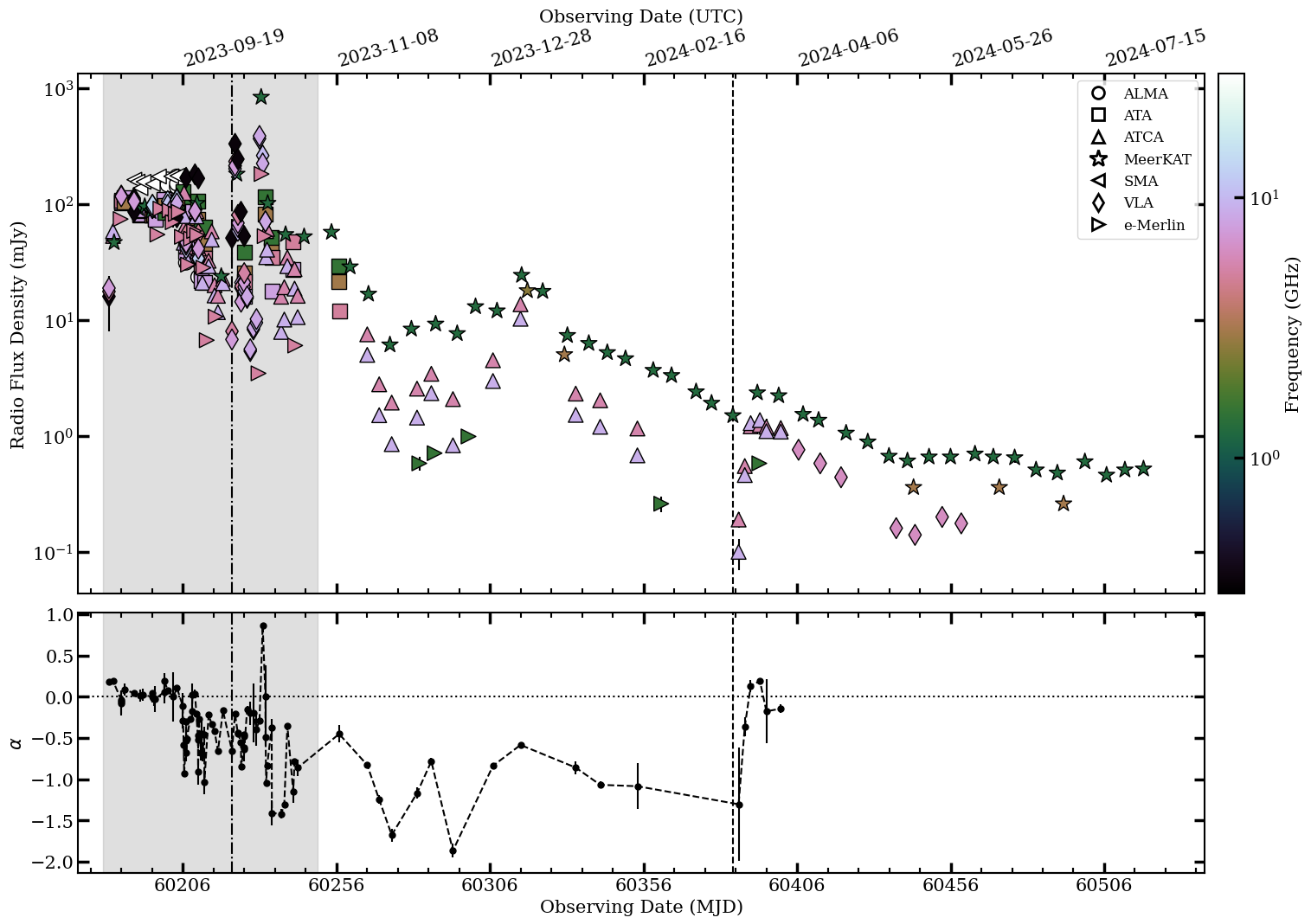}
            \caption{The full radio light curves of Swift J1727.8$-$1613 during its 2023--2024 outburst: (\textit{top panel}) the integrated flux density evolution. Each marker corresponds to a different telescope, and the marker colours correspond to the observing frequency. To ease visualization, we have saturated the colour axis at 30\,GHz; the saturation only affects observation taken with ALMA or the SMA. (\textit{bottom panel}) the inter-band spectral index evolution. The dashed-dotted and dashed lines correspond to the hard-to-soft (2024 October 5; MJD 60222) and soft-to-hard (2024 March 15; MJD 60385) state transitions, respectively. The grey-shaded region highlights the period of bright radio flaring that is shown in Figure~\ref{fig:zoom_radio_LC}.}            
            \label{fig:full_radio_LC}
    \end{figure}
\end{turnpage}
\twocolumngrid

\begin{figure*}
    \centering
    \includegraphics[width=1.0\linewidth]{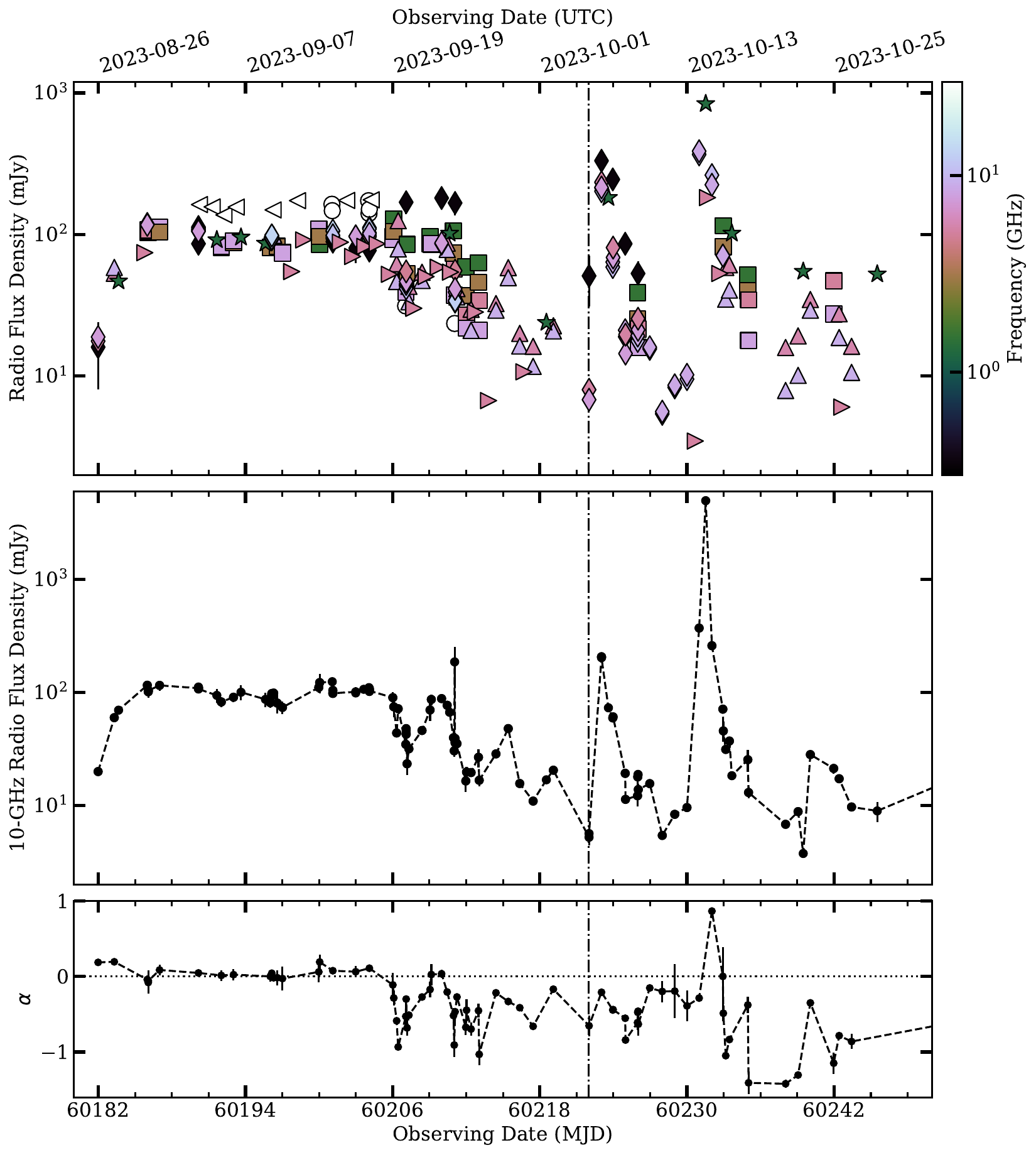}
    \caption{A zoomed-in view of the radio flaring: (\textit{top panel}) multi-frequency, multi-facility light curves where the markers follow the same scheme as Figure~\ref{fig:full_radio_LC}; (\textit{middle panel}) monochromatic radio light curves, where each of the measured flux densities have been scaled to 10\,GHz adopting the closest (in time) spectral index measurement (excluding high-frequency SMA and ALMA data); (\textit{bottom panel}) the inter-band spectral indices. The source exhibited multiple bright radio flares, suggesting multiple jet ejection events.}
    \label{fig:zoom_radio_LC}
\end{figure*}

\section{Results and Discussion} \label{sec:results}

This section presents the radio observations of the 2023--2024 outburst of Swift J1727. To highlight the simultaneous X-ray evolution, we include the X-ray hardness-intensity diagram (HID) using the publicly available data from the Monitor of All sky X-ray Image Gas Slit Camera \citep[MAXI/GSC;][]{MAXI2009,maxigsc}. The HID is presented in Figure~\ref{fig:XrayHID}, where the arrows show the time-evolution of the source and the marker colours trace the radio flux density (scaled to 10$\,$GHz following the routine discussed in Section~\ref{sec:HS_Trans}). The green diamond and yellow star correspond to the reported hard-to-soft \citep{Bollemeijer2023a,Bollemeijer2023b} and soft-to-hard \citep{Podgorny2024} state transitions, respectively. The increase in the hardness ratio following the (initial) reporting of a hard-to-soft state transition is consistent with the reporting by \citet{Wenfei2023} that stated the source returned to the hard state; as a result, Swift J1727 almost certainly underwent multiple hard-to-soft state transitions. From the HID, it is clear that Swift J1727 completed the canonical loop (or `turtle head') pattern seen from BH LMXB outbursts \citep[e.g.,][]{fender2004,belloni2010}, before fading below detectability, and likely returning to quiescence.

The remainder of this section focuses on the evolution of the flux densities and spectral indices from the jet. We only report statistical errors, but we note that telescopes have (frequency-dependent) systematic errors due to uncertainty in the absolute flux scale. Here, we present the integrated radio flux densities; for observations with multiple components (e.g., compact jet and jet ejecta or multiple jet ejecta), we sum the flux densities from the individual components, adding their errors in quadrature as an estimate of the integrated flux density errors. While in some cases, this will be summing information from the core and ejecta, which are physically decoupled, it mimics what would be measured with only low angular resolution monitoring and still captures much of the critical astrophysics. 

Aside from the image weightings mentioned in Section~\ref{sec:methods}, we do not apply any further manipulation of the visibilities during imaging (e.g., no $uv$-tapering). Care should be taken when combining data from different telescopes as they will probe different angular scales and, as a result, may resolve some flux density (this is primarily an issue for jet ejecta and when comparing e-Merlin to the other non-VLBI facilities). Regarding the radio spectral indices, we exclusively calculate inter-band values given that intra-band spectral indices have known biases \citep[e.g., as shown in the 1--2\,GHz VLA SDSS Stripe 82 survey;][]{Heywood2016}. As the source is highly variable, we only calculate these inter-band spectral indices for strictly simultaneous multi-frequency observations. By enforcing strict simultaneity, each spectral index is computed using flux densities from a single observatory (ATCA, VLA, or ATA), thereby avoiding complications arising from differences in angular scale sensitivity across telescopes. Moreover, the multi-frequency observations at these facilities are sensitive to angular scales much larger than the typical source size ($\lesssim$\,1''), mitigating the risk of artificial spectral features caused by frequency-dependent spatial filtering.

The radio properties are graphically represented in Figure~\ref{fig:full_radio_LC} and \ref{fig:zoom_radio_LC}. We include per-observatory light curves in Appendix A, and the raw data is presented in Appendix B. The data and plotting scripts are available via GitHub\footnote{\url{https://github.com/AKHughes1994/SwJ1727_2023_Outburst} with DOI: \dataset[10.5281/zenodo.15389817]{}}.

\subsection{Radio Outburst of Swift J1727.8$-$1613}
The radio evolution of Swift J1727 is shown in Figure~\ref{fig:full_radio_LC}: (top panel) lightcurves; (bottom panel) spectral indices. The dashed-dotted and dashed vertical lines correspond to the reported hard-to-soft (2024 October 5; MJD 60222) and soft-to-hard (2024 March 15; MJD 60385) state transitions, respectively \citep{Bollemeijer2023a,Bollemeijer2023b,Podgorny2024}.

\subsubsection{The Initial Hard State and State Transition(s)}
\label{sec:HS_Trans}
After the onset of the outburst (2023 August 24, MJD 60180), the source exhibited a sharp rise in its radio flux density, reaching $\sim$\,100\,mJy at all frequencies due to its flat spectral index (i.e., $\alpha\,{\sim}\,0$), consistent with emission from a partially self-absorbed compact jet. Swift J1727 remained at $\sim$\,100\,mJy for a few weeks before undergoing a decrease in radio flux density, followed by a period of flaring in the lead-up to (and immediately after) the reported hard-to-soft state transition. The radio flaring occurred (approximately) between 2023 September 20 (MJD 60207) and 2023 November 1 (MJD 60249), and the brightest flare reached a flux density of $\sim$\,1\,Jy at $\sim$\,1.28\,GHz. 

Figure~\ref{fig:zoom_radio_LC} shows a `zoomed-in' view of the flaring period (grey region, Fig.~\ref{fig:full_radio_LC}). The top panel adopts the same marker style and colour scheme as Figure~\ref{fig:full_radio_LC}. The middle panel presents a monochromatic light curve, where each observation is scaled to 10\ GHz using the nearest spectral index measurement. We fold the error in the spectral indexes into our monochromatic flux densities, adopting Gaussian error propagation. However, we note that spectral index variability is an unaccounted-for systematic effect on the scaled flux density of single-frequency observations (i.e., observations where an inter-band spectral index was not calculated). As a result, we exclude observations with central frequencies that have more than an order of magnitude separation from 10\,GHz (i.e., $<$\,1\,GHz or $>$\,100\,GHz) from the monochromatic light curves as large frequency separations will be more affected by errors in the adopted spectral indices. Moreover, systematic errors may significantly worsen during flaring when the radio spectral index is frequency-dependent and rapidly evolving. Despite the systematics, we include the monochromatic light curves as a qualitative description of the flaring progression, and we caution readers to view it as such. 

Both the poly and monochromatic light curves clearly show multiple radio flares consistent with the reporting from \citet{Wenfei2023} that \src\ underwent multiple hard-to-soft state transitions. The two most pronounced flares peaked at $\sim$\,300\,mJy (0.338\,GHz; MJD 60223) and $\sim$\,800\,mJy (1.28\,GHz; MJD 60231). Alongside the flaring, the radio spectral indices varied between $\alpha\,{\sim}$\,0.9 and $\alpha\,\,{\sim}\,-1.5$, on timescales as short as days. Simultaneous flux density and spectral evolution are typical of the flaring associated with the launching of jet ejecta \citep[e.g.,][]{fender2019}, and a decrease in flux density in the lead-up to flaring (attributed to compact jet quenching) has been observed in other BH LMXBs \citep[e.g.,][]{Brocksopp2002ejection,russel2019}. Recently, the launching of jet ejecta during flaring has been confirmed via direct imaging \citep{Wood2025SWJ1717Ejecta}.

Following the prescription from \citet{fender2019}, assuming equipartition of energy between particles and magnetic fields, and adopting distances of 3.7--5.5\,kpc \citep{Mata2025distance1727,Burridge2025distance1727}, we can estimate the minimum internal energy ($E_{\rm min}$) required to produce each flare. The observed flaring corresponds to an $E_{\rm min}\,{\sim}\,(0.6-2)\times10^{40}{\rm\,erg}$, consistent with the range of energies inferred from other sources \citep[albeit at the higher end;][]{fender2019,russell2019,bright2020,Carotenuto2021}. It should be noted that alternative and perhaps more robust methods of minimum energy estimation regularly arrive at values orders of magnitude higher than those inferred from flaring alone \citep[e.g., kinematic modelling or direct size measurements of ejecta;][]{bright2020,Carotenuto2022,Carotenuto2024}. As a result, these estimates are likely a conservative minimum. 

\subsubsection{The Soft State}
Following the flaring, Swift J1727 transitioned to the soft state, where its flux density began decreasing, reaching $<$\,10\,mJy at all observing frequencies. \src\ then underwent a modest re-brightening that peaked around 2024 January 14 (MJD 60322; $\sim$\,30\,mJy at 1.28\,GHz), followed by a continuation of the flux density decay. Throughout the soft state, the radio spectral index was optically thin $\alpha\,{\lesssim}-0.5$, consistent with the radio emission originating from jet ejecta, which is unsurprising as radio flaring is a known signature of jet ejection events \citep[e.g.,][]{Hjellming1995, Fender1999, rushton2017,millerjones2019,bright2020,Carotenuto2021} and compact jet emission is heavily quenched in the soft state \citep[e.g.,][]{Fender1999gx339quench,coriat2011,russell2019,russel2020,Maccarone2020quench,Carotenuto2021}. 

Interestingly, while optically thin, the spectral index was highly variable ($-2\,{\lesssim}\,\alpha\,{\lesssim}-0.5$), suggesting a significant evolution in the population of synchrotron emitting particles (likely electrons). The temporal evolution of the spectral index in the soft state is suggestive of particle (re-)acceleration possibly due to jet-ISM interactions \citep[e.g., the longest-lasting ejecta are thought to be driven to continued particle acceleration from jet-ISM interactions;][]{Corbel2002Ejecta,russell2019,bright2020,Carotenuto2021,Bahramian2023}. Indeed, the archetypal long-lived ejecta from XTE J1550$-$564 showed a similar transition from $\alpha\,{\sim}\,-1.2$ to $\alpha\,{\sim}\,-0.4$ which was interpreted as an interaction-driven renewal of particle acceleration \citep{Migliori2017XTE1550}.

For power-law distributed population synchrotron emitting electrons, $N(E)\text{d}E\propto E^{-p}\text{d}E$, an optically thin spectral index ($\alpha$) is related to the power-law index $p$ through $\alpha = -(p-1)/2$. Therefore, for typical values of $p\,{\sim}\,$2--3 \citep{longair2011}, optically thin synchrotron emission should have a spectral index between $-$0.5 and $-$1. Synchrotron cooling can result in more steep spectrum emission as $\alpha = -p/2$; however, spectral indexes of $\alpha\,{<}\,-1.5$ still require values of $p\,{>}\,3$ (e.g., the minimum of $\alpha = -1.86\,{\pm}\,0.08$ on MJD 60293). The origin of steep-spectrum radio sources remains an open question. Prior studies have suggested thermal (or quasi-thermal) particle distributions \citep{Pacholczyk1971Text}, ultra-relativistic shocks \citep{Ballard1992}, or energy exchange between the particles and turbulent magnetic fields \citep{Bell2019A} can produce steep spectra. However, most BH LMXBs showing comparably steep spectral indices \citep[i.e., $\alpha\,{\lesssim}\,-1.5;$][]{Russell2010XTESpectra,Shahbaz2013SteepSpectra} have typically been seen with infrared--optical observations of compact jets. Seeing such steep spectral indices for radio observations of ejecta provides an interesting case study for understanding particle acceleration from BH LMXB jets, especially through comparisons with steep-spectrum AGN \citep[e.g.,][]{Koljonen2015AGNXRBSpectra}.

\subsubsection{The Final Hard State and the Return to Quiescence}
Towards the end of the outburst, the radio emission of Swift J1727 re-brightened once more after returning to the hard state on (approximately) 2024 March 15 \citep[$\sim\,$MJD 60385;][]{Podgorny2024}. For example, ATCA detected an increase from ${\sim}\,0.1\,$mJy to ${\sim}\,1$\,mJy in both its 5.5 and 9.0\,GHz bands during observations taken within 4 days, consistent with a re-formation of the compact jet \citep[e.g.,][]{russell2014}. The spectral index evolved from $-$0.5 to $\sim$\,0, where the latter value is typical of a partially self-absorbed synchrotron radiation. The emission associated with the compact jet quickly decayed as the source returned to quiescence. Despite returning to X-ray quiescence, the radio emission plateaued due to long-lasting jet ejecta, presumably interacting with the ISM. Jet ejecta are typically optically thin, consistent with our observations having lower flux densities at higher frequencies. The slow decay, ejecta interactions, and, as a result, renewed particle acceleration make Swift J1727 an ideal candidate for investigating jet-based feedback.

\section{Conclusion} \label{sec:conclusions}
We have presented a comprehensive multi-frequency, multi-facility radio monitoring campaign of the BH LMXB Swift J1727 during its 2023--2024 outburst. The presented monitoring period began on 2023 August 25 (MJD 60181) and ended 2024 July 27 (MJD 60518), and includes 198 individual epochs with observing frequencies ranging between 0.3 to 225\ GHz, making Swift J1727 one of the most intensely monitored LMXBs to date. Our observations captured the complete outburst evolution, including both the hard-to-soft and soft-to-hard state transitions (i.e., the disruption and reformation of the compact jet) and, as a result, bright radio flares and spectral index variability. The primary motivation behind this work was to provide the community with comprehensive radio lightcurves; as a result, we only briefly (and qualitatively) interpreted the data, leaving more comprehensive targeted analyses for future publications by the comprising collaborations. We intend to continue rapidly releasing radio light curves following the cessation of future BH LMXB outbursts, as the radio properties will assist the interpretations drawn from monitoring programs at other wavelengths, akin to how the publicly available MAXI/GSC data is commonly used to interpret the X-ray and accretion state evolution for radio-centred analyses. We encourage other studies of this source to use these data where appropriate.

AKH thanks UKRI for support. AJT and EL acknowledge that this research was undertaken thanks to funding from the Canada Research Chairs Program and the support of the Natural Sciences and Engineering Research Council of Canada (NSERC; funding reference number RGPIN--2024--04458). FC acknowledges support from the Royal Society through the Newton International Fellowship programme (NIF/R1/211296). TDR is an INAF IAF fellow. RS acknowledges the INAF grant number 1.05.23.04.04. JvdE acknowledges a Warwick Astrophysics prize post-doctoral fellowship made possible thanks to a generous philanthropic donation, and was supported by funding from the European Union's Horizon Europe research and innovation programme under the Marie Sklodowska-Curie grant agreement No 101148693 (MeerSHOCKS) for part of this work. CMW acknowledges financial support from the Forrest Research Foundation Scholarship, the Jean-Pierre Macquart Scholarship, and the Australian Government Research Training Program Scholarship. This project has received funding from the European Research Council (ERC) under the European Union’s Horizon 2020 research and innovation programme (grant agreement No. 101002352, PI: M. Linares). 

This paper makes use of the following ALMA data: ADS/JAO.ALMA\#2022.1.01182.T. ALMA is a partnership of ESO (representing its member states), NSF (USA) and NINS (Japan), together with NRC (Canada), MOST and ASIAA (Taiwan), and KASI (Republic of Korea), in cooperation with the Republic of Chile. The Joint ALMA Observatory is operated by ESO, AUI/NRAO and NAOJ. The National Radio Astronomy Observatory is a facility of the National Science Foundation operated under cooperative agreement by Associated Universities, Inc. The Submillimeter Array is a joint project between the Smithsonian Astrophysical Observatory and the Academia Sinica Institute of Astronomy and Astrophysics and is funded by the Smithsonian Institution and the Academia Sinica. We recognize that Maunakea is a culturally important site for the indigenous Hawaiian people; we are privileged to study the cosmos from its summit. The MeerKAT telescope is operated by the South African Radio Astronomy Observatory, which is a facility of the National Research Foundation, an agency of the Department of Science and Innovation. We acknowledge the use of the Inter-University Institute for Data Intensive Astronomy (IDIA) data intensive research cloud for data processing. IDIA is a South African university partnership involving the University of Cape Town, the University of Pretoria and the University of the Western Cape. The Australia Telescope Compact Array is part of the Australia Telescope National Facility which is funded by the Australian Government for operation as a National Facility managed by CSIRO. We acknowledge the Gomeroi people as the traditional owners of the ATCA observatory site. \textit{e}-MERLIN is a National Facility operated by the University of Manchester at Jodrell Bank Observatory on behalf of STFC, part of UK Research and Innovation. This research has made use of the MAXI data provided by RIKEN, JAXA and the MAXI team.

\

\appendix
\section{Additional Figures}
\begin{figure*}[b]
    \centering
    \includegraphics[width=1.0\linewidth]{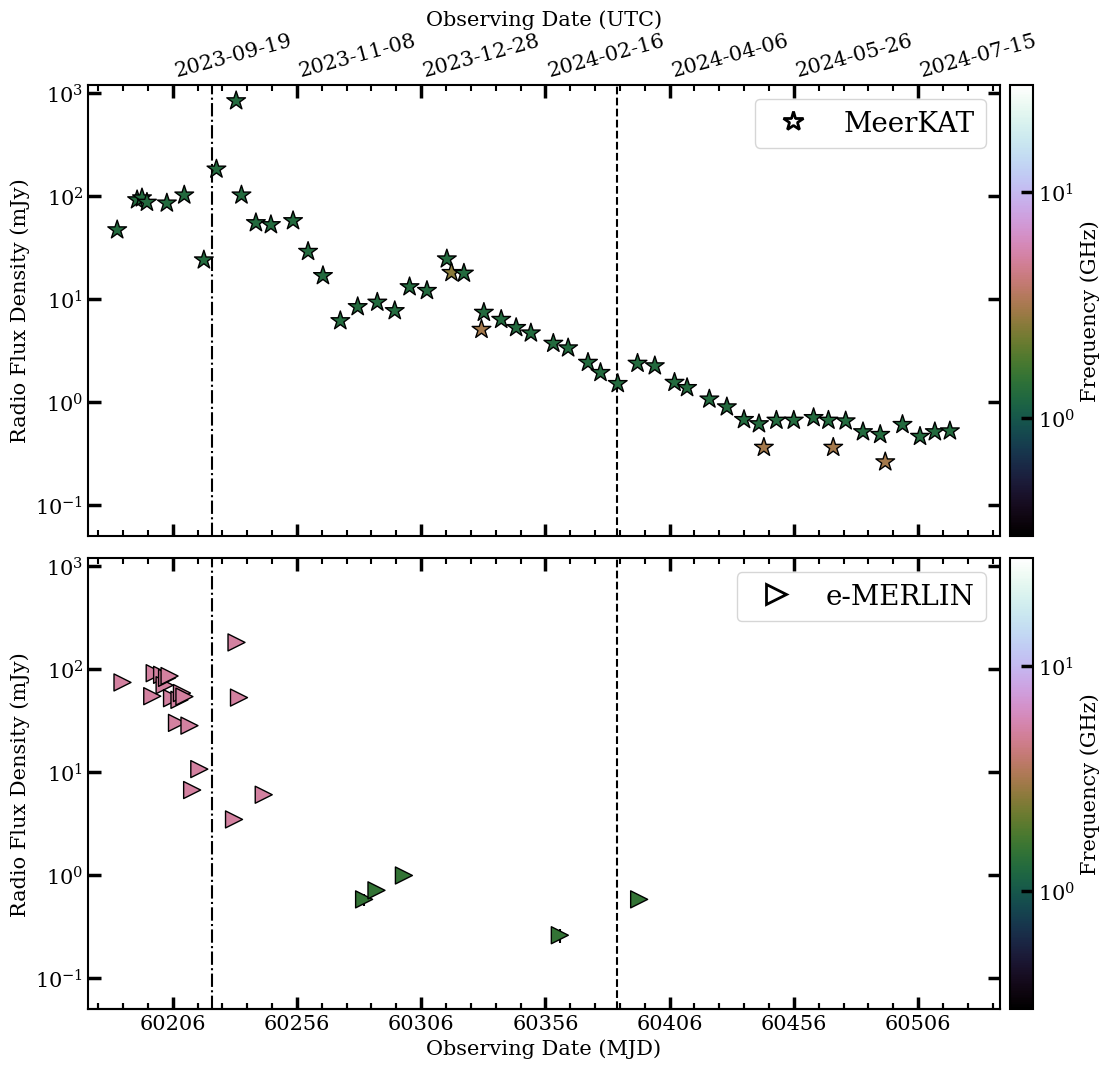}
    \caption{Radio light curves as seen by MeerKAT (\textit{top panel}) and e-MERLIN (\textit{bottom panel}). As in  Figure~\ref{fig:full_radio_LC}, the marker colours highlight the observing frequencies and the vertical lines mark the state transitions.}
    \label{fig:perobs1}
\end{figure*}

\begin{figure*}[t]
    \centering
    \includegraphics[width=1.0\linewidth]{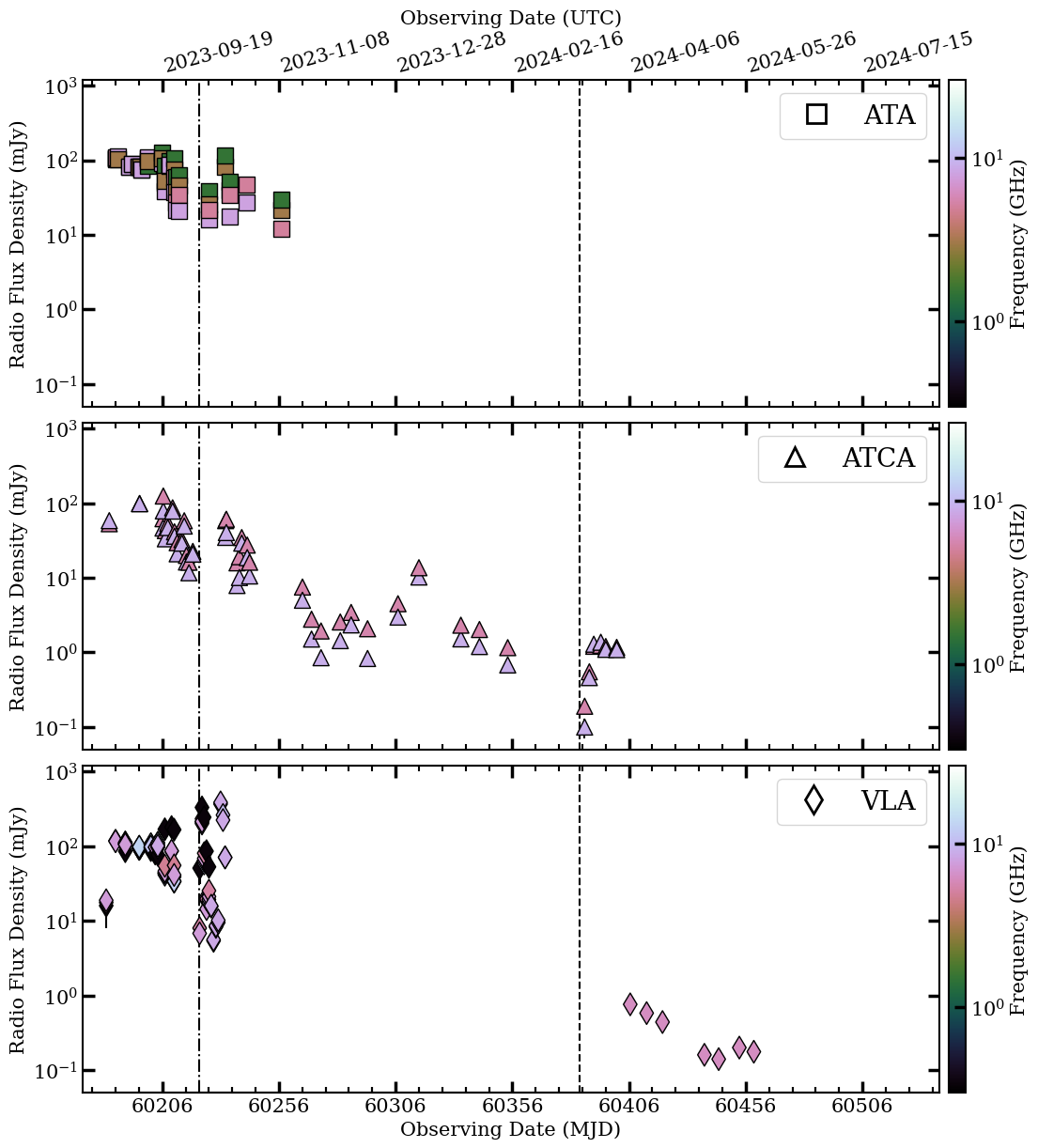}
    \caption{Same as Figure~\ref{fig:perobs1} but for ATA (\textit{top panel}), ATCA (\textit{middle panel}), and the VLA (\textit{bottom panel}).}
    \label{fig:perobs2}
\end{figure*}

\begin{figure*}[t]
    \centering
    \includegraphics[width=1.0\linewidth]{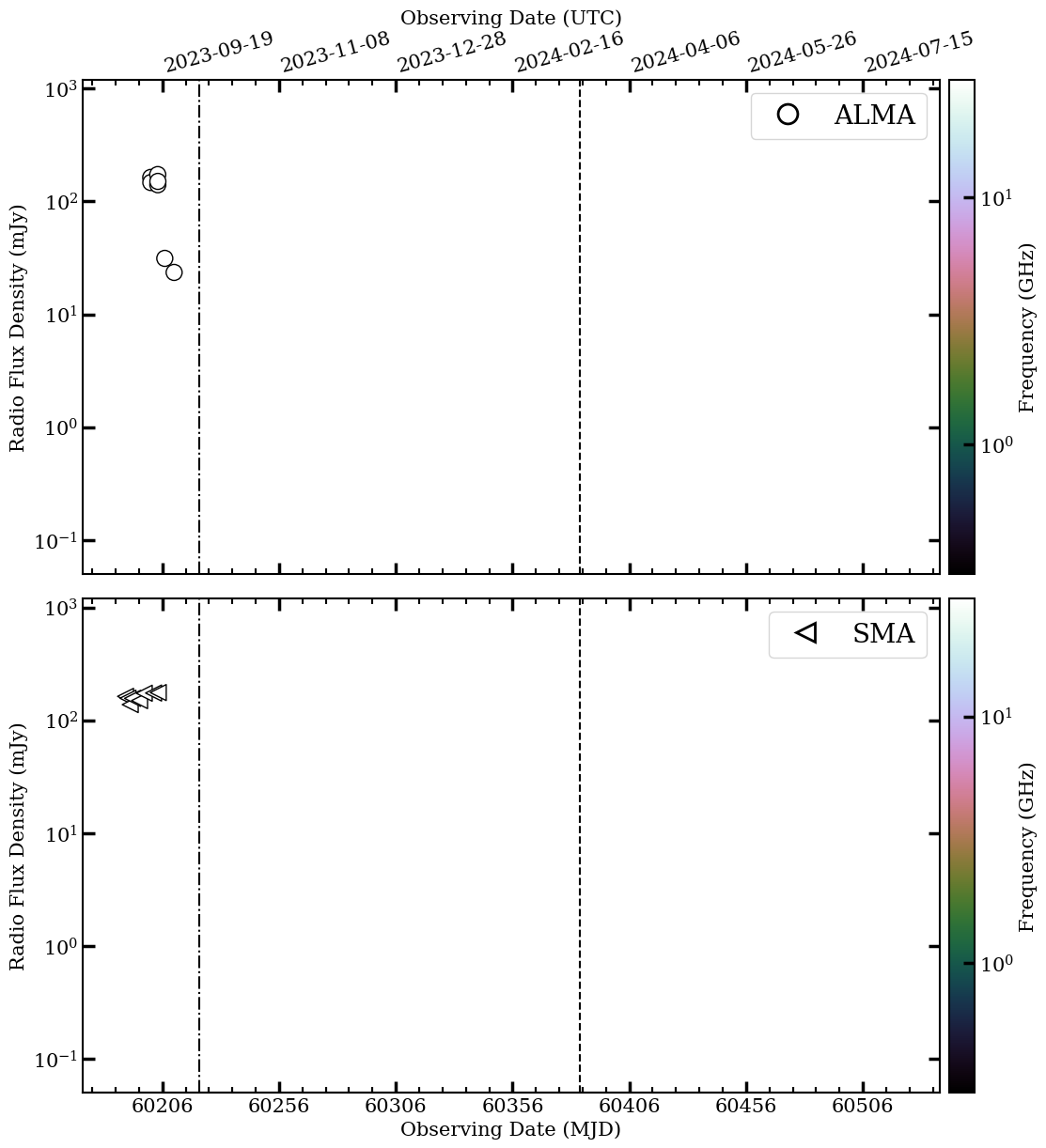}
    \caption{Same as Figure~\ref{fig:perobs1} but for ALMA (\textit{top panel}) and the SMA (\textit{bottom panel}).}
    \label{fig:perobs3}
\end{figure*}

Figures~\ref{fig:perobs1}, ~\ref{fig:perobs2}, and ~\ref{fig:perobs3} show per-observatory per-observatory light curves. 

\section{Data Tables}
Table~\ref{tab:radio_coverage} contains the radio flux densities and observing parameters (e.g., central frequency, duration, observing date) for each epoch.  
\onecolumngrid
\begin{turnpage}
    \begin{table}[htp]
        \centering
            \caption{Table containing each epoch's observation and flux density information. The variables $t_{\rm mid}$, $\Delta t$, $\nu_{\rm ctr}$, and $F_\nu$ correspond to the midpoint of the observation, approximate observation length, central frequency, and integrated flux density, respectively.}
            \begin{tabular}{rrrrrrrr}
            \Xhline{5\arrayrulewidth}
            \multicolumn{1}{c}{$t_{\rm mid}$~(UTC)} & \multicolumn{1}{c}{$t_{\rm mid}$~(MJD)} & \multicolumn{1}{c}{$\Delta t$~(min)} & \multicolumn{1}{c}{Telescope} & \multicolumn{1}{c}{$\nu_{\rm ctr}$~(GHz)} & \multicolumn{1}{c}{$F_{\nu}$~(mJy)} & \multicolumn{1}{c}{PI Name} & \multicolumn{1}{c}{Project ID}\\ 
            \Xhline{5\arrayrulewidth}
2023-08-25 23:35:48 & 60181.9832 & 20 & VLA & 0.338 & $16\,{\pm}\,8$ & Miller-Jones & VLITE/23A-260 \\
2023-08-25 23:35:48 & 60181.9832 & 20 & VLA & 5.2 & $17.64\,{\pm}\,0.03$ & Miller-Jones & 23A-260 \\
2023-08-25 23:35:48 & 60181.9832 & 20 & VLA & 7.5 & $18.88\,{\pm}\,0.04$ & Miller-Jones & 23A-260 \\
2023-08-27 07:13:43 & 60183.3012 & 210 & ATCA & 5.5 & $53.2\,{\pm}\,0.7$ & Russell/Carotenuto & C2601/C3057/C3362/CX550 \\
2023-08-27 07:13:43 & 60183.3012 & 210 & ATCA & 9.0 & $58.5\,{\pm}\,0.3$ & Russell/Carotenuto & C2601/C3057/C3362/CX550 \\
2023-08-27 15:35:25 & 60183.6496 & 15 & MeerKAT & 1.28 & $46.77\,{\pm}\,0.03$ & Fender & X-KAT/ThunderKAT \\
2023-08-29 18:59:02 & 60185.791 & 265 & e-MERLIN & 5.07 & $74.4\,{\pm}\,0.1$ & Williams-Baldwin/Carotenuto & RR16003 \\
2023-08-29 23:49:29 & 60185.9927 & 20 & VLA & 5.2 & $118.73\,{\pm}\,0.03$ & Miller-Jones & 23A-260 \\
2023-08-29 23:49:29 & 60185.9927 & 20 & VLA & 7.5 & $116.93\,{\pm}\,0.04$ & Miller-Jones & 23A-260 \\
2023-08-30 01:43:49 & 60186.0721 & 176 & ATA & 8.0 & $103\,{\pm}\,5$ & Bright & --- \\
2023-08-30 01:43:49 & 60186.0721 & 176 & ATA & 5.0 & $107\,{\pm}\,5$ & Bright & --- \\
2023-08-30 01:54:54 & 60186.0798 & 154 & ATA & 8.0 & $105\,{\pm}\,5$ & Bright & --- \\
2023-08-30 01:54:54 & 60186.0798 & 154 & ATA & 5.0 & $108\,{\pm}\,5$ & Bright & --- \\
2023-08-30 23:53:31 & 60186.9955 & 63 & ATA & 8.0 & $113\,{\pm}\,6$ & Bright & --- \\
2023-08-30 23:53:31 & 60186.9955 & 63 & ATA & 3.0 & $104\,{\pm}\,5$ & Bright & --- \\
2023-09-03 03:59:36 & 60190.1664 & 210 & VLA & 4.75 & $105.21\,{\pm}\,0.07$ & Tetarenko & 22B-069 \\
2023-09-03 03:59:36 & 60190.1664 & 210 & VLA & 0.338 & $86\,{\pm}\,13$ & Tetarenko & VLITE/22B-069 \\
2023-09-03 03:59:36 & 60190.1664 & 210 & VLA & 8.5 & $108.39\,{\pm}\,0.08$ & Tetarenko & 22B-069 \\
2023-09-03 03:59:36 & 60190.1664 & 210 & VLA & 13.5 & $112.9\,{\pm}\,0.2$ & Tetarenko & 22B-069 \\
2023-09-03 03:59:36 & 60190.1664 & 210 & VLA & 12.5 & $111.0\,{\pm}\,0.2$ & Tetarenko & 22B-069 \\
2023-09-03 03:59:36 & 60190.1664 & 210 & VLA & 11.0 & $107.92\,{\pm}\,0.09$ & Tetarenko & 22B-069 \\
2023-09-03 03:59:36 & 60190.1664 & 210 & VLA & 7.5 & $105.59\,{\pm}\,0.06$ & Tetarenko & 22B-069 \\
2023-09-03 05:54:48 & 60190.2464 & 148 & SMA & 225.0 & $162.6\,{\pm}\,0.4$ & Vrtilek/Gurwell & 2023A-S051 \\
2023-09-04 06:41:19 & 60191.2787 & 69 & SMA & 225.0 & $156.6\,{\pm}\,0.5$ & Vrtilek/Gurwell & 2023A-S051 \\
2023-09-04 15:55:43 & 60191.6637 & 15 & MeerKAT & 1.28 & $91.58\,{\pm}\,0.04$ & Fender & X-KAT/ThunderKAT \\
2023-09-05 00:42:54 & 60192.0298 & 104 & ATA & 3.0 & $81\,{\pm}\,4$ & Bright & --- \\
2023-09-05 00:42:54 & 60192.0298 & 104 & ATA & 8.0 & $82\,{\pm}\,4$ & Bright & --- \\
2023-09-05 05:50:38 & 60192.2435 & 188 & SMA & 225.0 & $137.5\,{\pm}\,0.5$ & Vrtilek/Gurwell & 2023A-S025 \\
2023-09-06 00:36:34 & 60193.0254 & 145 & ATA & 3.0 & $88\,{\pm}\,4$ & Bright & --- \\
2023-09-06 00:36:34 & 60193.0254 & 145 & ATA & 8.0 & $90\,{\pm}\,5$ & Bright & --- \\
2023-09-06 05:47:11 & 60193.2411 & 180 & SMA & 225.0 & $156.1\,{\pm}\,1.4$ & Vrtilek/Gurwell & 2023A-S025 \\
            \Xhline{5\arrayrulewidth}
            \label{tab:radio_coverage}
            \end{tabular}
    \end{table}
\end{turnpage}
\twocolumngrid

\newpage
\clearpage
\onecolumngrid
\begin{turnpage}
    \begin{table}[htp]
    \renewcommand\thetable{1} 
        \centering
            \caption{Continued.}
            \begin{tabular}{rrrrrrrr}
            \Xhline{5\arrayrulewidth}
            \multicolumn{1}{c}{$t_{\rm mid}$~(UTC)} & \multicolumn{1}{c}{$t_{\rm mid}$~(MJD)} & \multicolumn{1}{c}{$\Delta t$~(min)} & \multicolumn{1}{c}{Telescope} & \multicolumn{1}{c}{$\nu_{\rm ctr}$~(GHz)} & \multicolumn{1}{c}{$F_{\nu}$~(mJy)} & \multicolumn{1}{c}{PI Name} & \multicolumn{1}{c}{Project ID}\\ 
            \Xhline{5\arrayrulewidth}
2023-09-06 15:14:06 & 60193.6348 & 15 & MeerKAT & 1.28 & $95.57\,{\pm}\,0.05$ & Fender & X-KAT/ThunderKAT \\
2023-09-08 15:05:28 & 60195.6288 & 15 & MeerKAT & 1.28 & $86.64\,{\pm}\,0.04$ & Fender & X-KAT/ThunderKAT \\
2023-09-09 00:33:15 & 60196.0231 & 112 & ATA & 8.0 & $81\,{\pm}\,4$ & Bright & --- \\
2023-09-09 00:33:15 & 60196.0231 & 112 & ATA & 3.0 & $81\,{\pm}\,4$ & Bright & --- \\
2023-09-09 03:29:39 & 60196.1456 & 210 & VLA & 4.75 & $92.56\,{\pm}\,0.04$ & Tetarenko & 22B-069 \\
2023-09-09 03:29:39 & 60196.1456 & 210 & VLA & 0.338 & $93\,{\pm}\,15$ & Tetarenko & VLITE/22B-069 \\
2023-09-09 03:29:39 & 60196.1456 & 210 & VLA & 8.5 & $93.47\,{\pm}\,0.05$ & Tetarenko & 22B-069 \\
2023-09-09 03:29:39 & 60196.1456 & 210 & VLA & 7.5 & $92.53\,{\pm}\,0.05$ & Tetarenko & 22B-069 \\
2023-09-09 03:29:39 & 60196.1456 & 210 & VLA & 12.5 & $98.56\,{\pm}\,0.07$ & Tetarenko & 22B-069 \\
2023-09-09 03:29:39 & 60196.1456 & 210 & VLA & 11.0 & $93.87\,{\pm}\,0.06$ & Tetarenko & 22B-069 \\
2023-09-09 03:29:39 & 60196.1456 & 210 & VLA & 13.5 & $98.5\,{\pm}\,0.2$ & Tetarenko & 22B-069 \\
2023-09-09 06:02:52 & 60196.252 & 146 & SMA & 225.0 & $149.2\,{\pm}\,0.4$ & Vrtilek/Gurwell & 2023A-S025 \\
2023-09-09 07:06:31 & 60196.2962 & 324 & ATCA & 5.5 & $99.5\,{\pm}\,0.5$ & Russell/Carotenuto & C2601/C3057/C3362/CX550 \\
2023-09-09 07:06:31 & 60196.2962 & 324 & ATCA & 9.0 & $98.6\,{\pm}\,0.5$ & Russell/Carotenuto & C2601/C3057/C3362/CX550 \\
2023-09-09 14:13:12 & 60196.5925 & 1543 & ATA & 1.5 & $83\,{\pm}\,4$ & Bright & --- \\
2023-09-09 14:13:12 & 60196.5925 & 1543 & ATA & 3.0 & $82\,{\pm}\,4$ & Bright & --- \\
2023-09-10 00:24:11 & 60197.0168 & 71 & ATA & 5.0 & $75\,{\pm}\,4$ & Bright & --- \\
2023-09-10 00:24:11 & 60197.0168 & 71 & ATA & 8.0 & $74\,{\pm}\,4$ & Bright & --- \\
2023-09-10 18:24:02 & 60197.7667 & 330 & e-MERLIN & 5.07 & $54.7\,{\pm}\,0.1$ & Williams-Baldwin/Carotenuto & RR16003 \\
2023-09-11 05:50:38 & 60198.2435 & 140 & SMA & 225.0 & $173.4\,{\pm}\,0.5$ & Vrtilek/Gurwell & 2023A-S025 \\
2023-09-11 17:59:08 & 60198.7494 & 180 & e-MERLIN & 5.07 & $91.5\,{\pm}\,0.1$ & Williams-Baldwin/Carotenuto & RR16003 \\
2023-09-12 23:28:27 & 60199.9781 & 71 & ATA & 5.0 & $106\,{\pm}\,5$ & Bright & --- \\
2023-09-12 23:28:27 & 60199.9781 & 71 & ATA & 8.0 & $109\,{\pm}\,5$ & Bright & --- \\
2023-09-13 01:15:27 & 60200.0524 & 71 & ATA & 1.5 & $85\,{\pm}\,4$ & Bright & --- \\
2023-09-13 01:15:27 & 60200.0524 & 71 & ATA & 3.0 & $97\,{\pm}\,5$ & Bright & --- \\
2023-09-14 01:11:08 & 60201.0494 & 13 & ALMA & 145.0 & $163.5\,{\pm}\,0.04$ & Tetarenko & 2022.1.01182.T \\
2023-09-14 01:53:11 & 60201.0786 & 15 & ALMA & 97.5 & $147.3\,{\pm}\,0.04$ & Tetarenko & 2022.1.01182.T \\
2023-09-14 02:59:42 & 60201.1248 & 210 & VLA & 0.338 & $89\,{\pm}\,14$ & Tetarenko & VLITE/22B-069 \\
2023-09-14 02:59:42 & 60201.1248 & 210 & VLA & 12.5 & $105.2\,{\pm}\,0.3$ & Tetarenko & 22B-069 \\
2023-09-14 02:59:42 & 60201.1248 & 210 & VLA & 13.5 & $106.6\,{\pm}\,0.3$ & Tetarenko & 22B-069 \\
            \Xhline{5\arrayrulewidth}
            \end{tabular}
    \end{table}
\end{turnpage}
\twocolumngrid

\newpage
\clearpage
\onecolumngrid
\begin{turnpage}
    \begin{table}[htp]
    \renewcommand\thetable{1} 
        \centering
            \caption{Continued.}
            \begin{tabular}{rrrrrrrr}
            \Xhline{5\arrayrulewidth}
            \multicolumn{1}{c}{$t_{\rm mid}$~(UTC)} & \multicolumn{1}{c}{$t_{\rm mid}$~(MJD)} & \multicolumn{1}{c}{$\Delta t$~(min)} & \multicolumn{1}{c}{Telescope} & \multicolumn{1}{c}{$\nu_{\rm ctr}$~(GHz)} & \multicolumn{1}{c}{$F_{\nu}$~(mJy)} & \multicolumn{1}{c}{PI Name} & \multicolumn{1}{c}{Project ID}\\ 
            \Xhline{5\arrayrulewidth}
2023-09-14 02:59:42 & 60201.1248 & 210 & VLA & 4.75 & $93.31\,{\pm}\,0.03$ & Tetarenko & 22B-069 \\
2023-09-14 02:59:42 & 60201.1248 & 210 & VLA & 7.5 & $95.53\,{\pm}\,0.06$ & Tetarenko & 22B-069 \\
2023-09-14 02:59:42 & 60201.1248 & 210 & VLA & 8.5 & $97.93\,{\pm}\,0.07$ & Tetarenko & 22B-069 \\
2023-09-14 02:59:42 & 60201.1248 & 210 & VLA & 11.0 & $98.77\,{\pm}\,0.07$ & Tetarenko & 22B-069 \\
2023-09-14 17:47:02 & 60201.741 & 180 & e-MERLIN & 5.07 & $88.2\,{\pm}\,0.3$ & Williams-Baldwin/Carotenuto & RR16003 \\
2023-09-15 06:03:44 & 60202.2526 & 79 & SMA & 225.0 & $174\,{\pm}\,1$ & Vrtilek/Gurwell & 2023A-S024 \\
2023-09-15 17:43:00 & 60202.7382 & 180 & e-MERLIN & 5.07 & $69.9\,{\pm}\,0.1$ & Williams-Baldwin/Carotenuto & RR16003 \\
2023-09-15 23:47:11 & 60202.9911 & 20 & VLA & 0.338 & $80\,{\pm}\,17$ & Miller-Jones & VLITE/23A-260 \\
2023-09-15 23:47:11 & 60202.9911 & 20 & VLA & 5.2 & $97\,{\pm}\,8$ & Miller-Jones & 23A-260 \\
2023-09-15 23:47:11 & 60202.9911 & 20 & VLA & 7.5 & $97\,{\pm}\,7$ & Miller-Jones & 23A-260 \\
2023-09-16 15:30:40 & 60203.6463 & 15 & MeerKAT & 1.28 & $85.15\,{\pm}\,0.04$ & Fender & X-KAT/ThunderKAT \\
2023-09-16 17:38:58 & 60203.7354 & 180 & e-MERLIN & 5.07 & $82.4\,{\pm}\,0.1$ & Williams-Baldwin/Carotenuto & RR16003 \\
2023-09-17 00:57:36 & 60204.04 & 17 & ALMA & 233.0 & $173.2\,{\pm}\,0.05$ & Tetarenko & 2022.1.01182.T \\
2023-09-17 01:43:23 & 60204.0718 & 10 & ALMA & 97.5 & $140.8\,{\pm}\,0.03$ & Tetarenko & 2022.1.01182.T \\
2023-09-17 02:23:51 & 60204.0999 & 13 & ALMA & 145.0 & $150.7\,{\pm}\,0.03$ & Tetarenko & 2022.1.01182.T \\
2023-09-17 02:29:36 & 60204.1039 & 210 & VLA & 11.0 & $103.16\,{\pm}\,0.08$ & Tetarenko & 22B-069 \\
2023-09-17 02:29:36 & 60204.1039 & 210 & VLA & 0.338 & $77\,{\pm}\,12$ & Tetarenko & VLITE/22B-069 \\
2023-09-17 02:29:36 & 60204.1039 & 210 & VLA & 7.5 & $98.6\,{\pm}\,0.07$ & Tetarenko & 22B-069 \\
2023-09-17 02:29:36 & 60204.1039 & 210 & VLA & 4.75 & $96.28\,{\pm}\,0.08$ & Tetarenko & 22B-069 \\
2023-09-17 02:29:36 & 60204.1039 & 210 & VLA & 13.5 & $111.7\,{\pm}\,0.2$ & Tetarenko & 22B-069 \\
2023-09-17 02:29:36 & 60204.1039 & 210 & VLA & 12.5 & $109.3\,{\pm}\,0.2$ & Tetarenko & 22B-069 \\
2023-09-17 02:29:36 & 60204.1039 & 210 & VLA & 8.5 & $101.46\,{\pm}\,0.07$ & Tetarenko & 22B-069 \\
2023-09-17 06:00:08 & 60204.2501 & 99 & SMA & 225.0 & $175.8\,{\pm}\,0.4$ & Vrtilek/Gurwell & 2023A-S024 \\
2023-09-17 17:34:56 & 60204.7326 & 180 & e-MERLIN & 5.07 & $85.9\,{\pm}\,0.1$ & Williams-Baldwin/Carotenuto & RR16003 \\
2023-09-18 17:40:59 & 60205.7368 & 174 & e-MERLIN & 5.07 & $52.3\,{\pm}\,0.2$ & Williams-Baldwin/Carotenuto & RR16003 \\
2023-09-19 00:24:54 & 60206.0173 & 104 & ATA & 5.0 & $97\,{\pm}\,5$ & Bright & --- \\
2023-09-19 00:24:54 & 60206.0173 & 104 & ATA & 8.0 & $92\,{\pm}\,5$ & Bright & --- \\
2023-09-19 02:17:22 & 60206.0954 & 71 & ATA & 1.5 & $128\,{\pm}\,6$ & Bright & --- \\
2023-09-19 02:17:22 & 60206.0954 & 71 & ATA & 3.0 & $105\,{\pm}\,5$ & Bright & --- \\
2023-09-19 07:51:36 & 60206.3275 & 160 & ATCA & 5.5 & $62.1\,{\pm}\,0.3$ & Russell/Carotenuto & C2601/C3057/C3362/CX550 \\
            \Xhline{5\arrayrulewidth}
            \end{tabular}
    \end{table}
\end{turnpage}
\twocolumngrid

\newpage
\clearpage
\onecolumngrid
\begin{turnpage}
    \begin{table}[htp]
    \renewcommand\thetable{1} 
        \centering
            \caption{Continued.}
            \begin{tabular}{rrrrrrrr}
            \Xhline{5\arrayrulewidth}
            \multicolumn{1}{c}{$t_{\rm mid}$~(UTC)} & \multicolumn{1}{c}{$t_{\rm mid}$~(MJD)} & \multicolumn{1}{c}{$\Delta t$~(min)} & \multicolumn{1}{c}{Telescope} & \multicolumn{1}{c}{$\nu_{\rm ctr}$~(GHz)} & \multicolumn{1}{c}{$F_{\nu}$~(mJy)} & \multicolumn{1}{c}{PI Name} & \multicolumn{1}{c}{Project ID}\\ 
            \Xhline{5\arrayrulewidth}
2023-09-19 07:51:36 & 60206.3275 & 160 & ATCA & 9.0 & $46.5\,{\pm}\,0.5$ & Russell/Carotenuto & C2601/C3057/C3362/CX550 \\
2023-09-19 10:52:19 & 60206.453 & 133 & ATCA & 9.0 & $79.0\,{\pm}\,0.5$ & Russell/Carotenuto & C2601/C3057/C3362/CX550 \\
2023-09-19 10:52:19 & 60206.453 & 133 & ATCA & 5.5 & $125.1\,{\pm}\,0.7$ & Russell/Carotenuto & C2601/C3057/C3362/CX550 \\
2023-09-20 01:23:48 & 60207.0582 & 17 & ALMA & 233.0 & $31.3\,{\pm}\,0.04$ & Tetarenko & 2022.1.01182.T \\
2023-09-20 01:48:08 & 60207.0751 & 112 & ATA & 5.0 & $50\,{\pm}\,3$ & Bright & --- \\
2023-09-20 01:48:08 & 60207.0751 & 112 & ATA & 8.0 & $39\,{\pm}\,2$ & Bright & --- \\
2023-09-20 02:29:36 & 60207.1039 & 210 & VLA & 0.338 & $169\,{\pm}\,30$ & Tetarenko & VLITE/22B-069 \\
2023-09-20 02:29:36 & 60207.1039 & 210 & VLA & 13.5 & $43.7\,{\pm}\,0.4$ & Tetarenko & 22B-069 \\
2023-09-20 02:29:36 & 60207.1039 & 210 & VLA & 12.5 & $43.3\,{\pm}\,0.3$ & Tetarenko & 22B-069 \\
2023-09-20 02:29:36 & 60207.1039 & 210 & VLA & 11.0 & $41.6\,{\pm}\,0.3$ & Tetarenko & 22B-069 \\
2023-09-20 02:29:36 & 60207.1039 & 210 & VLA & 8.5 & $44.7\,{\pm}\,0.3$ & Tetarenko & 22B-069 \\
2023-09-20 02:29:36 & 60207.1039 & 210 & VLA & 7.5 & $46.5\,{\pm}\,0.1$ & Tetarenko & 22B-069 \\
2023-09-20 02:29:36 & 60207.1039 & 210 & VLA & 4.75 & $54.8\,{\pm}\,0.1$ & Tetarenko & 22B-069 \\
2023-09-20 04:15:53 & 60207.1777 & 112 & ATA & 1.5 & $85\,{\pm}\,4$ & Bright & --- \\
2023-09-20 04:15:53 & 60207.1777 & 112 & ATA & 3.0 & $53\,{\pm}\,3$ & Bright & --- \\
2023-09-20 07:51:10 & 60207.3272 & 372 & ATCA & 5.5 & $43.0\,{\pm}\,0.3$ & Russell/Carotenuto & C2601/C3057/C3362/CX550 \\
2023-09-20 07:51:10 & 60207.3272 & 372 & ATCA & 9.0 & $33.4\,{\pm}\,0.2$ & Russell/Carotenuto & C2601/C3057/C3362/CX550 \\
2023-09-20 17:25:00 & 60207.7257 & 168 & e-MERLIN & 5.07 & $30.1\,{\pm}\,0.1$ & Williams-Baldwin/Carotenuto & RR16003 \\
2023-09-21 09:33:07 & 60208.398 & 372 & ATCA & 5.5 & $54.1\,{\pm}\,0.6$ & Russell/Carotenuto & C2601/C3057/C3362/CX550 \\
2023-09-21 09:33:07 & 60208.398 & 372 & ATCA & 9.0 & $47.3\,{\pm}\,0.6$ & Russell/Carotenuto & C2601/C3057/C3362/CX550 \\
2023-09-21 17:22:59 & 60208.7243 & 180 & e-MERLIN & 5.07 & $50.4\,{\pm}\,0.2$ & Williams-Baldwin/Carotenuto & RR16003 \\
2023-09-22 01:16:36 & 60209.0532 & 112 & ATA & 1.5 & $97\,{\pm}\,5$ & Bright & --- \\
2023-09-22 01:16:36 & 60209.0532 & 112 & ATA & 3.0 & $86\,{\pm}\,4$ & Bright & --- \\
2023-09-22 03:44:29 & 60209.1559 & 112 & ATA & 5.0 & $85\,{\pm}\,4$ & Bright & --- \\
2023-09-22 03:44:29 & 60209.1559 & 112 & ATA & 8.0 & $86\,{\pm}\,4$ & Bright & --- \\
2023-09-22 17:18:57 & 60209.7215 & 180 & e-MERLIN & 5.07 & $58.7\,{\pm}\,0.1$ & Williams-Baldwin/Carotenuto & RR16003 \\
2023-09-22 23:50:21 & 60209.9933 & 20 & VLA & 0.338 & $181\,{\pm}\,30$ & Miller-Jones & VLITE/23A-260 \\
2023-09-22 23:51:04 & 60209.9938 & 20 & VLA & 5.2 & $86\,{\pm}\,1$ & Miller-Jones & 23A-260 \\
2023-09-22 23:51:04 & 60209.9938 & 20 & VLA & 7.5 & $87.1\,{\pm}\,1.4$ & Miller-Jones & 23A-260 \\
2023-09-23 10:45:33 & 60210.4483 & 190 & ATCA & 5.5 & $86.9\,{\pm}\,0.1$ & Russell/Carotenuto & C2601/C3057/C3362/CX550 \\
            \Xhline{5\arrayrulewidth}
            \end{tabular}
    \end{table}
\end{turnpage}
\twocolumngrid

\newpage
\clearpage
\onecolumngrid
\begin{turnpage}
    \begin{table}[htp]
    \renewcommand\thetable{1} 
        \centering
            \caption{Continued.}
            \begin{tabular}{rrrrrrrr}
            \Xhline{5\arrayrulewidth}
            \multicolumn{1}{c}{$t_{\rm mid}$~(UTC)} & \multicolumn{1}{c}{$t_{\rm mid}$~(MJD)} & \multicolumn{1}{c}{$\Delta t$~(min)} & \multicolumn{1}{c}{Telescope} & \multicolumn{1}{c}{$\nu_{\rm ctr}$~(GHz)} & \multicolumn{1}{c}{$F_{\nu}$~(mJy)} & \multicolumn{1}{c}{PI Name} & \multicolumn{1}{c}{Project ID}\\ 
            \Xhline{5\arrayrulewidth}
2023-09-23 10:45:33 & 60210.4483 & 190 & ATCA & 9.0 & $78.5\,{\pm}\,0.1$ & Russell/Carotenuto & C2601/C3057/C3362/CX550 \\
2023-09-23 15:30:31 & 60210.6462 & 15 & MeerKAT & 1.28 & $101.87\,{\pm}\,0.04$ & Fender & X-KAT/ThunderKAT \\
2023-09-23 17:18:57 & 60210.7215 & 180 & e-MERLIN & 5.07 & $54.2\,{\pm}\,0.1$ & Williams-Baldwin/Carotenuto & RR16003 \\
2023-09-23 23:07:52 & 60210.9638 & 51 & ATA & 1.5 & $106\,{\pm}\,5$ & Bright & --- \\
2023-09-23 23:07:52 & 60210.9638 & 51 & ATA & 3.0 & $74\,{\pm}\,4$ & Bright & --- \\
2023-09-24 00:34:50 & 60211.0242 & 51 & ATA & 5.0 & $57\,{\pm}\,3$ & Bright & --- \\
2023-09-24 00:34:50 & 60211.0242 & 51 & ATA & 8.0 & $37.2\,{\pm}\,1.9$ & Bright & --- \\
2023-09-24 01:24:14 & 60211.0585 & 10 & ALMA & 97.5 & $23.5\,{\pm}\,0.03$ & Tetarenko & 2022.1.01182.T \\
2023-09-24 02:14:38 & 60211.0935 & 210 & VLA & 4.75 & $55.5\,{\pm}\,0.2$ & Tetarenko & 22B-069 \\
2023-09-24 02:14:38 & 60211.0935 & 210 & VLA & 0.338 & $167\,{\pm}\,30$ & Tetarenko & VLITE/22B-069 \\
2023-09-24 02:14:38 & 60211.0935 & 210 & VLA & 8.5 & $39.1\,{\pm}\,0.3$ & Tetarenko & 22B-069 \\
2023-09-24 02:14:38 & 60211.0935 & 210 & VLA & 13.5 & $33.5\,{\pm}\,0.1$ & Tetarenko & 22B-069 \\
2023-09-24 02:14:38 & 60211.0935 & 210 & VLA & 11.0 & $34.1\,{\pm}\,0.2$ & Tetarenko & 22B-069 \\
2023-09-24 02:14:38 & 60211.0935 & 210 & VLA & 12.5 & $34.0\,{\pm}\,0.2$ & Tetarenko & 22B-069 \\
2023-09-24 02:14:38 & 60211.0935 & 210 & VLA & 7.5 & $41.3\,{\pm}\,0.1$ & Tetarenko & 22B-069 \\
2023-09-24 05:56:06 & 60211.2473 & 357 & ATCA & 5.5 & $41.4\,{\pm}\,0.4$ & Russell/Carotenuto & C2601/C3057/C3362/CX550 \\
2023-09-24 05:56:06 & 60211.2473 & 357 & ATCA & 9.0 & $36.2\,{\pm}\,0.5$ & Russell/Carotenuto & C2601/C3057/C3362/CX550 \\
2023-09-24 23:14:29 & 60211.9684 & 84 & ATA & 1.5 & $59\,{\pm}\,3$ & Bright & --- \\
2023-09-24 23:14:29 & 60211.9684 & 84 & ATA & 3.0 & $37.0\,{\pm}\,1.8$ & Bright & --- \\
2023-09-25 00:41:28 & 60212.0288 & 83 & ATA & 5.0 & $26.9\,{\pm}\,1.3$ & Bright & --- \\
2023-09-25 00:41:28 & 60212.0288 & 83 & ATA & 8.0 & $21.8\,{\pm}\,1.1$ & Bright & --- \\
2023-09-25 09:32:24 & 60212.3975 & 347 & ATCA & 5.5 & $29.6\,{\pm}\,0.8$ & Russell/Carotenuto & C2601/C3057/C3362/CX550 \\
2023-09-25 09:32:24 & 60212.3975 & 347 & ATCA & 9.0 & $21.0\,{\pm}\,0.7$ & Russell/Carotenuto & C2601/C3057/C3362/CX550 \\
2023-09-25 17:58:59 & 60212.7493 & 180 & e-MERLIN & 5.07 & $28.3\,{\pm}\,0.1$ & Williams-Baldwin/Carotenuto & RR16003 \\
2023-09-26 00:02:26 & 60213.0017 & 51 & ATA & 1.5 & $63\,{\pm}\,3$ & Bright & --- \\
2023-09-26 00:02:26 & 60213.0017 & 51 & ATA & 3.0 & $46\,{\pm}\,2$ & Bright & --- \\
2023-09-26 01:29:16 & 60213.062 & 51 & ATA & 5.0 & $34.1\,{\pm}\,1.7$ & Bright & --- \\
2023-09-26 01:29:16 & 60213.062 & 51 & ATA & 8.0 & $21\,{\pm}\,1$ & Bright & --- \\
2023-09-26 19:44:32 & 60213.8226 & 60 & e-MERLIN & 5.07 & $6.7\,{\pm}\,0.2$ & Williams-Baldwin/Carotenuto & RR16003 \\
2023-09-27 10:16:45 & 60214.4283 & 262 & ATCA & 5.5 & $32.5\,{\pm}\,0.2$ & Russell/Carotenuto & C2601/C3057/C3362/CX550 \\
            \Xhline{5\arrayrulewidth}
            \end{tabular}
    \end{table}
\end{turnpage}
\twocolumngrid

\newpage
\clearpage
\onecolumngrid
\begin{turnpage}
    \begin{table}[htp]
    \renewcommand\thetable{1} 
        \centering
            \caption{Continued.}
            \begin{tabular}{rrrrrrrr}
            \Xhline{5\arrayrulewidth}
            \multicolumn{1}{c}{$t_{\rm mid}$~(UTC)} & \multicolumn{1}{c}{$t_{\rm mid}$~(MJD)} & \multicolumn{1}{c}{$\Delta t$~(min)} & \multicolumn{1}{c}{Telescope} & \multicolumn{1}{c}{$\nu_{\rm ctr}$~(GHz)} & \multicolumn{1}{c}{$F_{\nu}$~(mJy)} & \multicolumn{1}{c}{PI Name} & \multicolumn{1}{c}{Project ID}\\ 
            \Xhline{5\arrayrulewidth}
2023-09-27 10:16:45 & 60214.4283 & 262 & ATCA & 9.0 & $29.2\,{\pm}\,0.5$ & Russell/Carotenuto & C2601/C3057/C3362/CX550 \\
2023-09-28 10:19:29 & 60215.4302 & 257 & ATCA & 5.5 & $58.3\,{\pm}\,0.4$ & Russell/Carotenuto & C2601/C3057/C3362/CX550 \\
2023-09-28 10:19:29 & 60215.4302 & 257 & ATCA & 9.0 & $49.5\,{\pm}\,0.6$ & Russell/Carotenuto & C2601/C3057/C3362/CX550 \\
2023-09-29 08:47:54 & 60216.3666 & 403 & ATCA & 5.5 & $20.0\,{\pm}\,0.01$ & Russell/Carotenuto & C2601/C3057/C3362/CX550 \\
2023-09-29 08:47:54 & 60216.3666 & 403 & ATCA & 9.0 & $16.31\,{\pm}\,0.01$ & Russell/Carotenuto & C2601/C3057/C3362/CX550 \\
2023-09-29 16:58:56 & 60216.7076 & 240 & e-MERLIN & 5.07 & $10.7\,{\pm}\,0.1$ & Williams-Baldwin/Carotenuto & RR16003 \\
2023-09-30 10:59:39 & 60217.4581 & 157 & ATCA & 5.5 & $16.2\,{\pm}\,0.1$ & Russell/Carotenuto & C2601/C3057/C3362/CX550 \\
2023-09-30 10:59:39 & 60217.4581 & 157 & ATCA & 9.0 & $11.7\,{\pm}\,0.1$ & Russell/Carotenuto & C2601/C3057/C3362/CX550 \\
2023-10-01 12:57:36 & 60218.54 & 15 & MeerKAT & 1.28 & $23.85\,{\pm}\,0.02$ & Fender & X-KAT/ThunderKAT \\
2023-10-02 02:35:31 & 60219.108 & 165 & ATCA & 5.5 & $22.64\,{\pm}\,0.02$ & Russell/Carotenuto & C2601/C3057/C3362/CX550 \\
2023-10-02 02:35:31 & 60219.108 & 165 & ATCA & 9.0 & $20.83\,{\pm}\,0.02$ & Russell/Carotenuto & C2601/C3057/C3362/CX550 \\
2023-10-05 00:27:47 & 60222.0193 & 30 & VLA & 0.338 & $51\,{\pm}\,20$ & Miller-Jones & VLITE/23A-260 \\
2023-10-05 00:27:47 & 60222.0193 & 30 & VLA & 5.2 & $8\,{\pm}\,1$ & Miller-Jones & 23A-260 \\
2023-10-05 00:27:47 & 60222.0193 & 30 & VLA & 7.5 & $6.8\,{\pm}\,1.1$ & Miller-Jones & 23A-260 \\
2023-10-06 00:49:58 & 60223.0347 & 30 & VLA & 8.5 & $213.57\,{\pm}\,0.06$ & Miller-Jones & 23A-260 \\
2023-10-06 00:49:58 & 60223.0347 & 30 & VLA & 11.0 & $202.19\,{\pm}\,0.09$ & Miller-Jones & 23A-260 \\
2023-10-06 00:49:58 & 60223.0347 & 30 & VLA & 5.2 & $233.7\,{\pm}\,1.1$ & Miller-Jones & 23A-260 \\
2023-10-06 00:49:58 & 60223.0347 & 30 & VLA & 7.5 & $215\,{\pm}\,1$ & Miller-Jones & 23A-260 \\
2023-10-06 00:49:58 & 60223.0347 & 30 & VLA & 0.338 & $332\,{\pm}\,50$ & Miller-Jones & VLITE/23A-260 \\
2023-10-06 14:21:50 & 60223.5985 & 15 & MeerKAT & 1.28 & $182.03\,{\pm}\,0.06$ & Fender & X-KAT/ThunderKAT \\
2023-10-06 22:54:02 & 60223.9542 & 30 & VLA & 10.2 & $58.86\,{\pm}\,0.07$ & Miller-Jones & 23A-260 \\
2023-10-06 22:54:02 & 60223.9542 & 30 & VLA & 8.5 & $63.84\,{\pm}\,0.06$ & Miller-Jones & 23A-260 \\
2023-10-06 22:54:02 & 60223.9542 & 30 & VLA & 0.338 & $245\,{\pm}\,40$ & Miller-Jones & VLITE/23A-260 \\
2023-10-06 23:29:54 & 60223.9791 & 30 & VLA & 7.5 & $68.8\,{\pm}\,0.4$ & Miller-Jones & 23A-260 \\
2023-10-06 23:29:54 & 60223.9791 & 30 & VLA & 5.2 & $80.8\,{\pm}\,0.3$ & Miller-Jones & 23A-260 \\
2023-10-07 23:13:55 & 60224.968 & 30 & VLA & 10.2 & $18.94\,{\pm}\,0.07$ & Miller-Jones & 23A-260 \\
2023-10-07 23:13:55 & 60224.968 & 30 & VLA & 8.5 & $21.07\,{\pm}\,0.06$ & Miller-Jones & 23A-260 \\
2023-10-07 23:13:55 & 60224.968 & 30 & VLA & 0.338 & $86\,{\pm}\,14$ & Miller-Jones & VLITE/23A-260 \\
2023-10-07 23:49:46 & 60224.9929 & 30 & VLA & 7.5 & $14.4\,{\pm}\,0.1$ & Miller-Jones & 23A-260 \\
2023-10-07 23:49:46 & 60224.9929 & 30 & VLA & 5.2 & $19.6\,{\pm}\,0.1$ & Miller-Jones & 23A-260 \\
            \Xhline{5\arrayrulewidth}
            \end{tabular}
    \end{table}
\end{turnpage}
\twocolumngrid

\newpage
\clearpage
\onecolumngrid
\begin{turnpage}
    \begin{table}[htp]
    \renewcommand\thetable{1} 
        \centering
            \caption{Continued.}
            \begin{tabular}{rrrrrrrr}
            \Xhline{5\arrayrulewidth}
            \multicolumn{1}{c}{$t_{\rm mid}$~(UTC)} & \multicolumn{1}{c}{$t_{\rm mid}$~(MJD)} & \multicolumn{1}{c}{$\Delta t$~(min)} & \multicolumn{1}{c}{Telescope} & \multicolumn{1}{c}{$\nu_{\rm ctr}$~(GHz)} & \multicolumn{1}{c}{$F_{\nu}$~(mJy)} & \multicolumn{1}{c}{PI Name} & \multicolumn{1}{c}{Project ID}\\ 
            \Xhline{5\arrayrulewidth}
2023-10-08 23:22:16 & 60225.9738 & 51 & ATA & 1.5 & $38.8\,{\pm}\,1.9$ & Bright & --- \\
2023-10-08 23:22:16 & 60225.9738 & 51 & ATA & 3.0 & $25.4\,{\pm}\,1.3$ & Bright & --- \\
2023-10-08 23:44:26 & 60225.9892 & 30 & VLA & 10.2 & $17.7\,{\pm}\,0.1$ & Miller-Jones & 23A-260 \\
2023-10-08 23:44:26 & 60225.9892 & 30 & VLA & 8.5 & $19.3\,{\pm}\,0.1$ & Miller-Jones & 23A-260 \\
2023-10-09 00:20:18 & 60226.0141 & 30 & VLA & 0.338 & $53\,{\pm}\,12$ & Miller-Jones & VLITE/23A-260 \\
2023-10-09 00:20:18 & 60226.0141 & 30 & VLA & 7.5 & $21.2\,{\pm}\,0.2$ & Miller-Jones & 23A-260 \\
2023-10-09 00:20:18 & 60226.0141 & 30 & VLA & 5.2 & $25.4\,{\pm}\,0.2$ & Miller-Jones & 23A-260 \\
2023-10-09 00:49:06 & 60226.0341 & 51 & ATA & 8.0 & $15.9\,{\pm}\,0.8$ & Bright & --- \\
2023-10-09 00:49:06 & 60226.0341 & 51 & ATA & 5.0 & $21.4\,{\pm}\,1.1$ & Bright & --- \\
2023-10-09 23:15:21 & 60226.969 & 30 & VLA & 10.2 & $15.53\,{\pm}\,0.08$ & Miller-Jones & 23A-260 \\
2023-10-09 23:15:21 & 60226.969 & 30 & VLA & 8.5 & $15.97\,{\pm}\,0.07$ & Miller-Jones & 23A-260 \\
2023-10-10 23:44:09 & 60227.989 & 30 & VLA & 10.2 & $5.4\,{\pm}\,0.1$ & Miller-Jones & 23A-260 \\
2023-10-10 23:44:09 & 60227.989 & 30 & VLA & 8.5 & $5.6\,{\pm}\,0.1$ & Miller-Jones & 23A-260 \\
2023-10-12 00:08:55 & 60229.0062 & 30 & VLA & 10.2 & $8.3\,{\pm}\,0.5$ & Miller-Jones & 23A-260 \\
2023-10-12 00:08:55 & 60229.0062 & 30 & VLA & 8.5 & $8.6\,{\pm}\,0.2$ & Miller-Jones & 23A-260 \\
2023-10-13 00:06:46 & 60230.0047 & 30 & VLA & 10.2 & $9.5\,{\pm}\,0.3$ & Miller-Jones & 23A-260 \\
2023-10-13 00:06:46 & 60230.0047 & 30 & VLA & 8.5 & $10.2\,{\pm}\,0.2$ & Miller-Jones & 23A-260 \\
2023-10-13 16:19:03 & 60230.6799 & 180 & e-MERLIN & 5.07 & $3.47\,{\pm}\,0.07$ & Williams-Baldwin/Carotenuto & RR16003 \\
2023-10-13 23:53:57 & 60230.9958 & 30 & VLA & 10.2 & $368.0\,{\pm}\,0.1$ & Miller-Jones & 23A-260 \\
2023-10-13 23:53:57 & 60230.9958 & 30 & VLA & 8.5 & $387.76\,{\pm}\,0.09$ & Miller-Jones & 23A-260 \\
2023-10-14 12:47:48 & 60231.5332 & 15 & MeerKAT & 1.28 & $839.4\,{\pm}\,0.3$ & Fender & X-KAT/ThunderKAT \\
2023-10-14 16:19:03 & 60231.6799 & 180 & e-MERLIN & 5.07 & $182.2\,{\pm}\,0.4$ & Williams-Baldwin/Carotenuto & RR16003 \\
2023-10-15 00:54:34 & 60232.0379 & 30 & VLA & 10.2 & $262.9\,{\pm}\,0.3$ & Miller-Jones & 23A-260 \\
2023-10-15 00:54:34 & 60232.0379 & 30 & VLA & 8.5 & $224.6\,{\pm}\,0.3$ & Miller-Jones & 23A-260 \\
2023-10-15 16:19:03 & 60232.6799 & 180 & e-MERLIN & 5.07 & $53.0\,{\pm}\,0.1$ & Williams-Baldwin/Carotenuto & RR16003 \\
2023-10-15 22:21:04 & 60232.9313 & 30 & VLA & 10.2 & $71\,{\pm}\,4$ & Miller-Jones & 23A-260 \\
2023-10-15 22:21:04 & 60232.9313 & 30 & VLA & 8.5 & $71\,{\pm}\,3$ & Miller-Jones & 23A-260 \\
2023-10-15 23:14:47 & 60232.9686 & 51 & ATA & 3.0 & $82\,{\pm}\,4$ & Bright & --- \\
2023-10-15 23:14:47 & 60232.9686 & 51 & ATA & 1.5 & $115\,{\pm}\,6$ & Bright & --- \\
2023-10-16 03:57:27 & 60233.1649 & 65 & ATCA & 9.0 & $34.9\,{\pm}\,0.1$ & Russell/Carotenuto & C2601/C3057/C3362/CX550 \\
            \Xhline{5\arrayrulewidth}
            \end{tabular}
    \end{table}
\end{turnpage}
\twocolumngrid

\newpage
\clearpage
\onecolumngrid
\begin{turnpage}
    \begin{table}[htp]
    \renewcommand\thetable{1} 
        \centering
            \caption{Continued.}
            \begin{tabular}{rrrrrrrr}
            \Xhline{5\arrayrulewidth}
            \multicolumn{1}{c}{$t_{\rm mid}$~(UTC)} & \multicolumn{1}{c}{$t_{\rm mid}$~(MJD)} & \multicolumn{1}{c}{$\Delta t$~(min)} & \multicolumn{1}{c}{Telescope} & \multicolumn{1}{c}{$\nu_{\rm ctr}$~(GHz)} & \multicolumn{1}{c}{$F_{\nu}$~(mJy)} & \multicolumn{1}{c}{PI Name} & \multicolumn{1}{c}{Project ID}\\ 
            \Xhline{5\arrayrulewidth}
2023-10-16 03:57:27 & 60233.1649 & 65 & ATCA & 5.5 & $58.5\,{\pm}\,0.1$ & Russell/Carotenuto & C2601/C3057/C3362/CX550 \\
2023-10-16 11:07:26 & 60233.4635 & 50 & ATCA & 5.5 & $61.1\,{\pm}\,0.1$ & Russell/Carotenuto & C2601/C3057/C3362/CX550 \\
2023-10-16 11:07:26 & 60233.4635 & 50 & ATCA & 9.0 & $40.5\,{\pm}\,0.1$ & Russell/Carotenuto & C2601/C3057/C3362/CX550 \\
2023-10-16 15:57:44 & 60233.6651 & 15 & MeerKAT & 1.28 & $101.9\,{\pm}\,0.05$ & Fender & X-KAT/ThunderKAT \\
2023-10-17 23:12:20 & 60234.9669 & 51 & ATA & 3.0 & $40\,{\pm}\,2$ & Bright & --- \\
2023-10-17 23:12:20 & 60234.9669 & 51 & ATA & 1.5 & $52\,{\pm}\,3$ & Bright & --- \\
2023-10-18 00:39:10 & 60235.0272 & 51 & ATA & 8.0 & $17.8\,{\pm}\,0.9$ & Bright & --- \\
2023-10-18 00:39:10 & 60235.0272 & 51 & ATA & 5.0 & $34.5\,{\pm}\,1.7$ & Bright & --- \\
2023-10-21 01:06:40 & 60238.0463 & 44 & ATCA & 9.0 & $7.9\,{\pm}\,0.2$ & Russell/Carotenuto & C2601/C3057/C3362/CX550 \\
2023-10-21 01:06:40 & 60238.0463 & 44 & ATCA & 5.5 & $15.9\,{\pm}\,0.2$ & Russell/Carotenuto & C2601/C3057/C3362/CX550 \\
2023-10-22 01:35:19 & 60239.0662 & 46 & ATCA & 9.0 & $10.1\,{\pm}\,0.1$ & Russell/Carotenuto & C2601/C3057/C3362/CX550 \\
2023-10-22 01:35:19 & 60239.0662 & 46 & ATCA & 5.5 & $19.2\,{\pm}\,0.2$ & Russell/Carotenuto & C2601/C3057/C3362/CX550 \\
2023-10-22 11:39:33 & 60239.4858 & 15 & MeerKAT & 1.28 & $54.84\,{\pm}\,0.03$ & Fender & X-KAT/ThunderKAT \\
2023-10-23 01:35:11 & 60240.0661 & 48 & ATCA & 5.5 & $34.7\,{\pm}\,0.1$ & Russell/Carotenuto & C2601/C3057/C3362/CX550 \\
2023-10-23 01:35:11 & 60240.0661 & 48 & ATCA & 9.0 & $29.2\,{\pm}\,0.1$ & Russell/Carotenuto & C2601/C3057/C3362/CX550 \\
2023-10-24 23:16:30 & 60241.9698 & 71 & ATA & 5.0 & $47\,{\pm}\,2$ & Bright & --- \\
2023-10-24 23:16:30 & 60241.9698 & 71 & ATA & 8.0 & $27.4\,{\pm}\,1.4$ & Bright & --- \\
2023-10-25 09:55:52 & 60242.4138 & 122 & ATCA & 5.5 & $27.5\,{\pm}\,0.1$ & Russell/Carotenuto & C2601/C3057/C3362/CX550 \\
2023-10-25 09:55:52 & 60242.4138 & 122 & ATCA & 9.0 & $18.69\,{\pm}\,0.07$ & Russell/Carotenuto & C2601/C3057/C3362/CX550 \\
2023-10-25 15:11:39 & 60242.6331 & 212 & e-MERLIN & 5.07 & $6.03\,{\pm}\,0.07$ & Williams-Baldwin/Carotenuto & RR16003 \\
2023-10-26 10:15:53 & 60243.4277 & 55 & ATCA & 9.0 & $10.6\,{\pm}\,0.5$ & Russell/Carotenuto & C2601/C3057/C3362/CX550 \\
2023-10-26 10:15:53 & 60243.4277 & 55 & ATCA & 5.5 & $16.2\,{\pm}\,0.2$ & Russell/Carotenuto & C2601/C3057/C3362/CX550 \\
2023-10-28 12:30:14 & 60245.521 & 15 & MeerKAT & 1.28 & $52.48\,{\pm}\,0.03$ & Fender & X-KAT/ThunderKAT \\
2023-11-06 10:29:25 & 60254.4371 & 15 & MeerKAT & 1.28 & $57.36\,{\pm}\,0.03$ & Fender & X-KAT/ThunderKAT \\
2023-11-08 22:04:39 & 60256.9199 & 71 & ATA & 3.0 & $21.5\,{\pm}\,1.1$ & Bright & --- \\
2023-11-08 22:04:39 & 60256.9199 & 71 & ATA & 1.5 & $29.4\,{\pm}\,1.5$ & Bright & --- \\
2023-11-08 23:51:12 & 60256.9939 & 71 & ATA & 5.0 & $11.9\,{\pm}\,0.6$ & Bright & --- \\
2023-11-12 11:33:30 & 60260.4816 & 15 & MeerKAT & 1.28 & $28.85\,{\pm}\,0.02$ & Fender & X-KAT/ThunderKAT \\
2023-11-18 01:49:43 & 60266.0762 & 76 & ATCA & 5.5 & $7.53\,{\pm}\,0.03$ & Russell/Carotenuto & C2601/C3057/C3362/CX550 \\
2023-11-18 01:49:43 & 60266.0762 & 76 & ATCA & 9.0 & $5.01\,{\pm}\,0.02$ & Russell/Carotenuto & C2601/C3057/C3362/CX550 \\
            \Xhline{5\arrayrulewidth}
            \end{tabular}
    \end{table}
\end{turnpage}
\twocolumngrid

\newpage
\clearpage
\onecolumngrid
\begin{turnpage}
    \begin{table}[htp]
    \renewcommand\thetable{1} 
        \centering
            \caption{Continued.}
            \begin{tabular}{rrrrrrrr}
            \Xhline{5\arrayrulewidth}
            \multicolumn{1}{c}{$t_{\rm mid}$~(UTC)} & \multicolumn{1}{c}{$t_{\rm mid}$~(MJD)} & \multicolumn{1}{c}{$\Delta t$~(min)} & \multicolumn{1}{c}{Telescope} & \multicolumn{1}{c}{$\nu_{\rm ctr}$~(GHz)} & \multicolumn{1}{c}{$F_{\nu}$~(mJy)} & \multicolumn{1}{c}{PI Name} & \multicolumn{1}{c}{Project ID}\\ 
            \Xhline{5\arrayrulewidth}
2023-11-18 11:19:06 & 60266.4716 & 15 & MeerKAT & 1.28 & $16.78\,{\pm}\,0.02$ & Fender & X-KAT/ThunderKAT \\
2023-11-21 23:09:27 & 60269.9649 & 132 & ATCA & 9.0 & $1.51\,{\pm}\,0.03$ & Russell/Carotenuto & C2601/C3057/C3362/CX550 \\
2023-11-21 23:09:27 & 60269.9649 & 132 & ATCA & 5.5 & $2.79\,{\pm}\,0.05$ & Russell/Carotenuto & C2601/C3057/C3362/CX550 \\
2023-11-25 11:02:41 & 60273.4602 & 15 & MeerKAT & 1.28 & $6.12\,{\pm}\,0.02$ & Fender & X-KAT/ThunderKAT \\
2023-11-26 00:55:17 & 60274.0384 & 188 & ATCA & 9.0 & $0.85\,{\pm}\,0.03$ & Russell/Carotenuto & C2601/C3057/C3362/CX550 \\
2023-11-26 00:55:17 & 60274.0384 & 188 & ATCA & 5.5 & $1.94\,{\pm}\,0.03$ & Russell/Carotenuto & C2601/C3057/C3362/CX550 \\
2023-12-02 10:48:25 & 60280.4503 & 15 & MeerKAT & 1.28 & $8.38\,{\pm}\,0.02$ & Fender & X-KAT/ThunderKAT \\
2023-12-04 06:48:05 & 60282.2834 & 183 & ATCA & 5.5 & $2.56\,{\pm}\,0.05$ & Russell/Carotenuto & C2601/C3057/C3362/CX550 \\
2023-12-04 06:48:05 & 60282.2834 & 183 & ATCA & 9.0 & $1.44\,{\pm}\,0.04$ & Russell/Carotenuto & C2601/C3057/C3362/CX550 \\
2023-12-05 00:53:34 & 60283.0372 & 336 & e-MERLIN & 1.51 & $0.58\,{\pm}\,0.08$ & Williams-Baldwin/Carotenuto & CY16208 \\
2023-12-08 22:29:25 & 60286.9371 & 183 & ATCA & 5.5 & $3.44\,{\pm}\,0.04$ & Russell/Carotenuto & C2601/C3057/C3362/CX550 \\
2023-12-08 22:29:25 & 60286.9371 & 183 & ATCA & 9.0 & $2.34\,{\pm}\,0.04$ & Russell/Carotenuto & C2601/C3057/C3362/CX550 \\
2023-12-10 00:23:54 & 60288.0166 & 420 & e-MERLIN & 1.51 & $0.71\,{\pm}\,0.05$ & Williams-Baldwin/Carotenuto & CY16208 \\
2023-12-10 08:10:36 & 60288.3407 & 15 & MeerKAT & 1.28 & $9.26\,{\pm}\,0.02$ & Fender & X-KAT/ThunderKAT \\
2023-12-15 23:25:52 & 60293.9763 & 292 & ATCA & 9.0 & $0.83\,{\pm}\,0.03$ & Russell/Carotenuto & C2601/C3057/C3362/CX550 \\
2023-12-15 23:25:52 & 60293.9763 & 292 & ATCA & 5.5 & $2.08\,{\pm}\,0.04$ & Russell/Carotenuto & C2601/C3057/C3362/CX550 \\
2023-12-17 08:44:52 & 60295.3645 & 15 & MeerKAT & 1.28 & $7.66\,{\pm}\,0.02$ & Fender & X-KAT/ThunderKAT \\
2023-12-21 01:18:28 & 60299.0545 & 497 & e-MERLIN & 1.51 & $0.99\,{\pm}\,0.07$ & Williams-Baldwin/Carotenuto & CY16208 \\
2023-12-23 07:50:35 & 60301.3268 & 15 & MeerKAT & 1.28 & $13.08\,{\pm}\,0.02$ & Fender & X-KAT/ThunderKAT \\
2023-12-29 00:09:30 & 60307.0066 & 327 & ATCA & 5.5 & $4.5\,{\pm}\,0.04$ & Russell/Carotenuto & C2601/C3057/C3362/CX550 \\
2023-12-29 00:09:30 & 60307.0066 & 327 & ATCA & 9.0 & $2.98\,{\pm}\,0.04$ & Russell/Carotenuto & C2601/C3057/C3362/CX550 \\
2023-12-30 07:47:08 & 60308.3244 & 15 & MeerKAT & 1.28 & $12.03\,{\pm}\,0.02$ & Fender & X-KAT/ThunderKAT \\
2024-01-06 23:27:01 & 60315.9771 & 237 & ATCA & 9.0 & $10.26\,{\pm}\,0.06$ & Russell/Carotenuto & C2601/C3057/C3362/CX550 \\
2024-01-06 23:27:01 & 60315.9771 & 237 & ATCA & 5.5 & $13.68\,{\pm}\,0.06$ & Russell/Carotenuto & C2601/C3057/C3362/CX550 \\
2024-01-07 07:58:04 & 60316.332 & 15 & MeerKAT & 1.28 & $24.39\,{\pm}\,0.02$ & Fender & X-KAT/ThunderKAT \\
2024-01-09 04:59:57 & 60318.2083 & 15 & MeerKAT & 2.62 & $17.98\,{\pm}\,0.01$ & Fender & X-KAT/ThunderKAT \\
2024-01-14 04:54:37 & 60323.2046 & 15 & MeerKAT & 1.28 & $17.72\,{\pm}\,0.02$ & Fender & X-KAT/ThunderKAT \\
2024-01-21 06:34:07 & 60330.2737 & 15 & MeerKAT & 3.06 & $5.04\,{\pm}\,0.02$ & Fender & X-KAT/ThunderKAT \\
2024-01-22 05:38:49 & 60331.2353 & 15 & MeerKAT & 1.28 & $7.38\,{\pm}\,0.03$ & Fender & X-KAT/ThunderKAT \\
2024-01-24 22:51:18 & 60333.9523 & 451 & ATCA & 9.0 & $1.52\,{\pm}\,0.05$ & Russell/Carotenuto & C2601/C3057/C3362/CX550 \\
            \Xhline{5\arrayrulewidth}
            \end{tabular}
    \end{table}
\end{turnpage}
\twocolumngrid

\newpage
\clearpage
\onecolumngrid
\begin{turnpage}
    \begin{table}[htp]
    \renewcommand\thetable{1} 
        \centering
            \caption{Continued.}
            \begin{tabular}{rrrrrrrr}
            \Xhline{5\arrayrulewidth}
            \multicolumn{1}{c}{$t_{\rm mid}$~(UTC)} & \multicolumn{1}{c}{$t_{\rm mid}$~(MJD)} & \multicolumn{1}{c}{$\Delta t$~(min)} & \multicolumn{1}{c}{Telescope} & \multicolumn{1}{c}{$\nu_{\rm ctr}$~(GHz)} & \multicolumn{1}{c}{$F_{\nu}$~(mJy)} & \multicolumn{1}{c}{PI Name} & \multicolumn{1}{c}{Project ID}\\ 
            \Xhline{5\arrayrulewidth}
2024-01-24 22:51:18 & 60333.9523 & 451 & ATCA & 5.5 & $2.32\,{\pm}\,0.05$ & Russell/Carotenuto & C2601/C3057/C3362/CX550 \\
2024-01-29 05:05:16 & 60338.212 & 15 & MeerKAT & 1.28 & $6.3\,{\pm}\,0.03$ & Fender & X-KAT/ThunderKAT \\
2024-02-01 21:53:08 & 60341.9119 & 367 & ATCA & 5.5 & $2.03\,{\pm}\,0.02$ & Russell/Carotenuto & C2601/C3057/C3362/CX550 \\
2024-02-01 21:53:08 & 60341.9119 & 367 & ATCA & 9.0 & $1.2\,{\pm}\,0.02$ & Russell/Carotenuto & C2601/C3057/C3362/CX550 \\
2024-02-04 05:33:47 & 60344.2318 & 15 & MeerKAT & 1.28 & $5.25\,{\pm}\,0.03$ & Fender & X-KAT/ThunderKAT \\
2024-02-10 03:46:13 & 60350.1571 & 15 & MeerKAT & 1.28 & $4.65\,{\pm}\,0.03$ & Fender & X-KAT/ThunderKAT \\
2024-02-14 00:44:47 & 60354.0311 & 253 & ATCA & 5.5 & $1.16\,{\pm}\,0.08$ & Russell/Carotenuto & C2601/C3057/C3362/CX550 \\
2024-02-14 00:44:47 & 60354.0311 & 253 & ATCA & 9.0 & $0.68\,{\pm}\,0.08$ & Russell/Carotenuto & C2601/C3057/C3362/CX550 \\
2024-02-19 04:29:16 & 60359.187 & 15 & MeerKAT & 1.28 & $3.69\,{\pm}\,0.03$ & Fender & X-KAT/ThunderKAT \\
2024-02-21 19:28:33 & 60361.8115 & 588 & e-MERLIN & 1.51 & $0.26\,{\pm}\,0.04$ & Williams-Baldwin/Carotenuto & CY16208 \\
2024-02-25 03:30:05 & 60365.1459 & 15 & MeerKAT & 1.28 & $3.33\,{\pm}\,0.03$ & Fender & X-KAT/ThunderKAT \\
2024-03-04 03:22:10 & 60373.1404 & 15 & MeerKAT & 1.28 & $2.41\,{\pm}\,0.03$ & Fender & X-KAT/ThunderKAT \\
2024-03-09 04:09:24 & 60378.1732 & 15 & MeerKAT & 1.28 & $1.92\,{\pm}\,0.02$ & Fender & X-KAT/ThunderKAT \\
2024-03-16 02:43:52 & 60385.1138 & 15 & MeerKAT & 1.28 & $1.5\,{\pm}\,0.03$ & Fender & X-KAT/ThunderKAT \\
2024-03-17 23:34:30 & 60386.9823 & 187 & ATCA & 5.5 & $0.19\,{\pm}\,0.03$ & Russell/Carotenuto & C2601/C3057/C3362/CX550 \\
2024-03-17 23:34:30 & 60386.9823 & 187 & ATCA & 9.0 & $0.1\,{\pm}\,0.03$ & Russell/Carotenuto & C2601/C3057/C3362/CX550 \\
2024-03-19 23:48:11 & 60388.9918 & 160 & ATCA & 5.5 & $0.55\,{\pm}\,0.02$ & Russell/Carotenuto & C2601/C3057/C3362/CX550 \\
2024-03-19 23:48:11 & 60388.9918 & 160 & ATCA & 9.0 & $0.46\,{\pm}\,0.02$ & Russell/Carotenuto & C2601/C3057/C3362/CX550 \\
2024-03-21 21:45:21 & 60390.9065 & 287 & ATCA & 5.5 & $1.21\,{\pm}\,0.03$ & Russell/Carotenuto & C2601/C3057/C3362/CX550 \\
2024-03-21 21:45:21 & 60390.9065 & 287 & ATCA & 9.0 & $1.29\,{\pm}\,0.03$ & Russell/Carotenuto & C2601/C3057/C3362/CX550 \\
2024-03-24 01:59:05 & 60393.0827 & 15 & MeerKAT & 1.28 & $2.36\,{\pm}\,0.03$ & Fender & X-KAT/ThunderKAT \\
2024-03-24 17:38:32 & 60393.7351 & 574 & e-MERLIN & 1.51 & $0.58\,{\pm}\,0.04$ & Williams-Baldwin/Carotenuto & CY16208 \\
2024-03-24 21:54:25 & 60393.9128 & 78 & ATCA & 5.5 & $1.246\,{\pm}\,0.003$ & Russell/Carotenuto & C2601/C3057/C3362/CX550 \\
2024-03-24 21:54:25 & 60393.9128 & 78 & ATCA & 9.0 & $1.37\,{\pm}\,0.01$ & Russell/Carotenuto & C2601/C3057/C3362/CX550 \\
2024-03-27 00:45:38 & 60396.0317 & 8 & ATCA & 5.5 & $1.2\,{\pm}\,0.2$ & Russell/Carotenuto & C2601/C3057/C3362/CX550 \\
2024-03-27 00:45:38 & 60396.0317 & 8 & ATCA & 9.0 & $1.1\,{\pm}\,0.1$ & Russell/Carotenuto & C2601/C3057/C3362/CX550 \\
2024-03-31 00:52:16 & 60400.0363 & 15 & MeerKAT & 1.28 & $2.23\,{\pm}\,0.03$ & Fender & X-KAT/ThunderKAT \\
2024-03-31 17:09:27 & 60400.7149 & 398 & ATCA & 5.5 & $1.17\,{\pm}\,0.02$ & Russell/Carotenuto & C2601/C3057/C3362/CX550 \\
2024-03-31 17:09:27 & 60400.7149 & 398 & ATCA & 9.0 & $1.09\,{\pm}\,0.02$ & Russell/Carotenuto & C2601/C3057/C3362/CX550 \\
2024-04-06 12:07:03 & 60406.5049 & 12 & VLA & 6.2 & $0.76\,{\pm}\,0.01$ & Plotkin & 23B-064 \\

            \Xhline{5\arrayrulewidth}
            \end{tabular}
    \end{table}
\end{turnpage}
\twocolumngrid

\newpage
\clearpage
\onecolumngrid
\begin{turnpage}
    \begin{table}[htp]
    \renewcommand\thetable{1} 
        \centering
            \caption{Continued.}
            \begin{tabular}{rrrrrrrr}
            \Xhline{5\arrayrulewidth}
            \multicolumn{1}{c}{$t_{\rm mid}$~(UTC)} & \multicolumn{1}{c}{$t_{\rm mid}$~(MJD)} & \multicolumn{1}{c}{$\Delta t$~(min)} & \multicolumn{1}{c}{Telescope} & \multicolumn{1}{c}{$\nu_{\rm ctr}$~(GHz)} & \multicolumn{1}{c}{$F_{\nu}$~(mJy)} & \multicolumn{1}{c}{PI Name} & \multicolumn{1}{c}{Project ID}\\ 
            \Xhline{5\arrayrulewidth}
2024-04-07 23:19:06 & 60407.9716 & 15 & MeerKAT & 1.28 & $1.54\,{\pm}\,0.03$ & Fender & X-KAT/ThunderKAT \\
2024-04-13 00:31:06 & 60413.0216 & 15 & MeerKAT & 1.28 & $1.37\,{\pm}\,0.03$ & Fender & X-KAT/ThunderKAT \\
2024-04-13 13:00:11 & 60413.5418 & 12 & VLA & 6.2 & $0.58\,{\pm}\,0.01$ & Plotkin & 23B-064 \\
2024-04-20 08:02:32 & 60420.3351 & 12 & VLA & 6.2 & $0.44\,{\pm}\,0.01$ & Plotkin & 23B-064 \\
2024-04-21 23:38:06 & 60421.9848 & 15 & MeerKAT & 1.28 & $1.06\,{\pm}\,0.03$ & Fender & X-KAT/ThunderKAT \\
2024-04-29 00:46:48 & 60429.0325 & 15 & MeerKAT & 1.28 & $0.89\,{\pm}\,0.03$ & Fender & X-KAT/ThunderKAT \\
2024-05-05 22:29:42 & 60435.9373 & 15 & MeerKAT & 1.28 & $0.67\,{\pm}\,0.03$ & Fender & X-KAT/ThunderKAT \\
2024-05-08 07:06:40 & 60438.2963 & 12 & VLA & 6.2 & $0.16\,{\pm}\,0.01$ & Plotkin & 23B-064 \\
2024-05-11 22:51:36 & 60441.9525 & 15 & MeerKAT & 1.28 & $0.61\,{\pm}\,0.03$ & Fender & X-KAT/ThunderKAT \\
2024-05-13 22:24:23 & 60443.9336 & 15 & MeerKAT & 3.06 & $0.36\,{\pm}\,0.02$ & Fender & X-KAT/ThunderKAT \\
2024-05-14 09:58:01 & 60444.4153 & 27 & VLA & 6.2 & $0.14\,{\pm}\,0.01$ & Plotkin & 23B-064 \\
2024-05-18 23:26:52 & 60448.977 & 15 & MeerKAT & 1.28 & $0.66\,{\pm}\,0.03$ & Fender & X-KAT/ThunderKAT \\
2024-05-23 06:04:19 & 60453.253 & 26 & VLA & 6.2 & $0.2\,{\pm}\,0.01$ & Plotkin & 23B-064 \\
2024-05-25 23:16:04 & 60455.9695 & 15 & MeerKAT & 1.28 & $0.66\,{\pm}\,0.03$ & Fender & X-KAT/ThunderKAT \\
2024-05-29 11:14:21 & 60459.4683 & 40 & VLA & 6.2 & $0.176\,{\pm}\,0.007$ & Plotkin & 23B-064 \\
2024-06-02 22:32:18 & 60463.9391 & 15 & MeerKAT & 1.28 & $0.7\,{\pm}\,0.03$ & Fender & X-KAT/ThunderKAT \\
2024-06-08 21:54:43 & 60469.913 & 30 & MeerKAT & 1.28 & $0.66\,{\pm}\,0.03$ & Fender & X-KAT/ThunderKAT \\
2024-06-10 20:21:07 & 60471.848 & 30 & MeerKAT & 3.06 & $0.36\,{\pm}\,0.01$ & Fender & X-KAT/ThunderKAT \\
2024-06-15 20:27:01 & 60476.8521 & 15 & MeerKAT & 1.28 & $0.65\,{\pm}\,0.06$ & Fender & X-KAT/ThunderKAT \\
2024-06-22 19:46:50 & 60483.8242 & 15 & MeerKAT & 1.28 & $0.51\,{\pm}\,0.03$ & Fender & X-KAT/ThunderKAT \\
2024-06-29 18:06:54 & 60490.7548 & 15 & MeerKAT & 1.28 & $0.48\,{\pm}\,0.03$ & Fender & X-KAT/ThunderKAT \\
2024-07-01 18:46:22 & 60492.7822 & 15 & MeerKAT & 3.06 & $0.26\,{\pm}\,0.01$ & Fender & X-KAT/ThunderKAT \\
2024-07-08 17:14:47 & 60499.7186 & 15 & MeerKAT & 1.28 & $0.6\,{\pm}\,0.03$ & Fender & X-KAT/ThunderKAT \\
2024-07-15 18:33:33 & 60506.7733 & 15 & MeerKAT & 1.28 & $0.46\,{\pm}\,0.03$ & Fender & X-KAT/ThunderKAT \\
2024-07-21 17:46:27 & 60512.7406 & 15 & MeerKAT & 1.28 & $0.51\,{\pm}\,0.03$ & Fender & X-KAT/ThunderKAT \\
2024-07-27 19:14:52 & 60518.802 & 15 & MeerKAT & 1.28 & $0.52\,{\pm}\,0.04$ & Fender & X-KAT/ThunderKAT \\
            \Xhline{5\arrayrulewidth}
            \end{tabular}
    \end{table}
\end{turnpage}
\twocolumngrid

\newpage
\clearpage
\bibliography{bibly}{}
\bibliographystyle{aasjournal}



\end{document}